\documentclass[12pt]{article}
\usepackage{amsmath}
\usepackage{graphicx}
\usepackage{enumerate}
\usepackage{natbib}

\newcommand{\blind}{1}

\usepackage[latin1]{inputenc}
\usepackage{amssymb, amsthm, tikz, csquotes}
\usepackage{amsmath, amssymb, amsthm, natbib, graphicx, tikz, xfrac, csquotes}
\usepackage[capposition=top]{floatrow}
\usepackage[colorlinks=true, linkcolor=blue, citecolor=blue]{hyperref}
\usetikzlibrary{decorations.pathreplacing} 
\newtheorem{assumption}{Assumption}
\newtheorem{lemma}{Lemma}
\newtheorem{theorem}{Theorem}

\newcommand{\dd}{\,\mathrm{d}}
\newcommand{\ol}{\overline}

\newcommand{\Dlim}{\overset{d}{\longrightarrow}}

\newcommand{\rT}{\lfloor r T \rfloor}

\newcommand{\sT}{\lfloor s T \rfloor}

\newcommand{\lvec}{\left(\begin{matrix}}
\newcommand{\rvec}{\end{matrix}\right)}

\usepackage[onehalfspacing]{setspace}
\newcommand{\supplement}{1}

\begin{document}

\def\spacingset#1{\renewcommand{\baselinestretch}%
{#1}\small\normalsize} \spacingset{1}


\if1\blind
{
  \title{Unit Root Testing with Slowly Varying Trends}
  \author{Sven Otto\thanks{Correspondence to: Sven Otto, University of Bonn, Institute for Finance and Statistics, Adenauerallee 24, 53113 Bonn, Germany. Tel.: +49-228-73-9271. Mail: sven.otto@.uni-bonn.de.}}
  \maketitle
} \fi

\if0\blind
{
  \bigskip
  \bigskip
  \bigskip
  \begin{center}
    {\LARGE\bf Unit Root Testing with Slowly Varying Trends}
\end{center}
  \medskip
} \fi

\bigskip
\begin{abstract}
\noindent
A unit root test is proposed for time series with a general nonlinear deterministic trend component.
It is shown that asymptotically the pooled OLS estimator of overlapping blocks filters out any trend component that satisfies some Lipschitz condition.
Under both fixed-$b$ and small-$b$ block asymptotics, the limiting distribution of the $t$-statistic for the unit root hypothesis is derived.
Nuisance parameter corrections provide heteroskedasticity-robust tests, and serial correlation is accounted for by pre-whitening.
A Monte Carlo study that considers slowly varying trends yields both good size and improved power results for the proposed tests when compared to conventional unit root tests.
\end{abstract}

\noindent%
 \vspace{1ex} 
\vfill
\newpage
\spacingset{1.5} 

\section{Introduction}

It is widely debated in the time series literature whether macroeconomic variables such as GDP, inflation, and interest rates are $I(1)$ or $I(0)$ around a deterministic trend.
Dickey-Fuller-type unit root tests often fail to reject the null hypothesis for these time series.
The trend component of a time series $y_t$ is typically treated as known up to some parameter vector.
The most commonly applied unit root tests, such as those developed by \cite{dickey1979}, \cite{said1984}, \cite{phillips1987}, \cite{phillipsperron1988}, and \cite{elliott1996}, impose either a constant or a linear trend model.
If, however, the deterministic trend component is nonlinear, highly persistent trend-stationary processes can be hardly distinguishable from unit root processes (see, e.g., \citealt{bierens1997} and \citealt{becker2006}).

It is not only a misspecified trend model that may lead to high power losses, as an overparameterized model can also reduce the power of unit root tests.
Therefore, many authors have suggested applying trend models that seem more suitable for macro data.
Broken trend models with one-time changes in mean or slope with known breakpoint were first studied by \cite{perron1989} and \cite{rappoport1989}.
\cite{christiano1992} demonstrated that a broken trend model with an unknown breakpoint is more adequate, and \cite{zivot1992}, as well as \cite{banerjee1992}, proposed unit root tests for this framework.
Structural changes in innovation variances were studied by \cite{hamori1997}, \cite{kim2002}, and \cite{cavaliere2005}, while \cite{cavaliere2011} considered unit root testing under broken trends together with nonstationary volatility.
\cite{leybourne1998}, \cite{kapetanios2003}, and \cite{kilic2011} allowed for exponential smooth transitions from one trend regime to another.
\cite{bierens1997} approximated a nonlinear mean function with Chebyshev polynomials, and \cite{enders2012a} proposed a Fourier series approximation of the trend, which are approaches that can be used when the exact form and date of structural changes are unknown.
For a comprehensive review on the research on unit root testing see \cite{choi2015}.

Dickey-Fuller-type tests are based on the $t$-statistic of the first-order autoregressive parameter.
In case of a constant trend, the estimator is derived from a regression of $\Delta y_t$ on $(y_{t-1} - \ol y)$, where $\ol y$ is the sample mean.
\cite{schmidt1992} estimated the constant by the initial observation, which results in a regression of $\Delta y_t$ on $(y_{t-1} - y_1)$.
Whereas a constant is often not a good global approximation, in a small block, a smoothly varying trend can be approximated quite closely by a constant. 
To exploit this fact, we propose a block procedure to filter out the unknown trend component.
Blocking was also used in \cite{rooch2019} to estimate the fractional integration parameter in a similar situation.
We divide the series into $T-B$ overlapping blocks of length $B$. 
As the blocks can be considered as units of a panel, we follow the panel unit root tests proposed by \cite{breitung2000} and \cite{levin2002} and consider a pooled regression of $\Delta y_{j+t}$ on $(y_{j+t-1} - y_j)$ for $2 \leq t\leq T$ and $1 \leq j \leq T-B$.
The deterministic function is approximated locally by a constant.
One could also use higher order local approximations of the trend function, but unreported simulations indicate that these approximations do not work well in samples of usual size.
For this reason, we focus on constant local approximations.
Under a general class of piecewise continuous trend functions, the resulting pooled estimator is consistent as $B,T \to \infty$.
The limiting null distribution of the t-statistic is a functional of a Brownian motion under fixed-$b$ asymptotics.
Under small-$b$ asymptotics, a normal distribution is obtained.

The paper is organized as follows:
In Section \ref{sec:model} the autoregressive model with independent and heteroskedastic errors is analyzed together with the asymptotic behavior of the pooled least squares estimator in the presence of a general nonlinear trend component.
For both fixed-$b$ and small-$b$ block asymptotics, the limiting distributions are derived under both the unit root hypothesis and under local alternatives.
In the presence of heteroskedastic errors, nuisance parameters appear in the limiting distributions, and the estimation of these parameters is discussed.
Section \ref{sec:testing} considers pseudo $t$-tests for the unit root hypothesis, and heteroskedasticity-robust test statistics are provided.
In Section \ref{sec_serialcorrelation}, a pre-whitening procedure is proposed in order to account for short-run dynamics, while Section \ref{sec_simulations} reports on Monte Carlo simulations. 
The tests are found to have only minor size distortions in small samples and are sized correctly in larger samples.
It is shown that in the presence of slowly varying trends, pooled tests tend to yield higher power than conventional unit root tests.
Finally, Section \ref{sec_conclusion1} presents the conclusion.

\if0\supplement
{
	While some proofs including those of the main theorems are presented in the Appendix, the more technical proofs are available as online supporting information.
}\fi
	In the following, $W(r)$ denotes a standard Brownian motion and ``$\Rightarrow$'' stands for weak convergence on the càdlàg space $D[0,1]$ together with a suitable norm.
$\Theta(\cdot)$ denotes the exact order Landau symbol, that is, $a_T = \Theta(b_T)$ if and only if $a_T=O(b_T)$ and $b_T = O(a_T)$, as $T \to \infty$.
Moreover, $\lfloor \cdot \rfloor$ is the integer part of its argument, and $\Delta y_t$ stands for the differenced series $y_t - y_{t-1}$. 
Finally, $\Dlim$ and $\overset{p}{\longrightarrow}$ denote convergence in distribution and convergence in probability.

\section{The pooled estimator} \label{sec:model}

We are interested in inference concerning the autoregressive parameter $\rho$ in the model
\begin{align}
	y_t = d_t + x_t, \quad x_t = \rho x_{t-1} + u_t, \quad t=1, \ldots, T,	\label{eq:model}
\end{align}
where $\rho$ is close or equal to one.
The deterministic trend component $d_t$ is treated as nonstochastic and fixed in repeated samples, where its functional form is nonparametric and unknown.

\begin{assumption}[trend component] \label{ass:trend}
The trend component is given by $d_t = d(t/T)$, where $d(r)$ is a piecewise Lipschitz continuous function.
\end{assumption}

Note that any continuously differentiable function is Lipschitz continuous.
Lipschitz functions are locally close to a constant value in the sense that there exists some $C < \infty$ such that $|d(r) - d(s)| \leq C|r-s|$ for all  $r,s \in \mathbb{R}$.
The piecewise Lipschitz condition allows for a partition with a finite number of intervals, such that $d(r)$ is Lipschitz continuous on each interval. 
This includes both smooth changes as well as abrupt breaks in the trend function.
For the initial value, it is assumed that $E[x_0^2] < \infty$.
We introduce the pooled estimator and the unit root test statistics under the following assumptions on the error term:

\begin{assumption}[heteroskedastic errors] \label{ass:errors}
The process $\{u_t\}_{t \in \mathbb{N}}$ is independently dis\-trib\-uted with $E[u_t]= 0$, $E[u_t^2] = \sigma_t^2$ and $E[u_t^4] < \infty$, where $\sigma_t = \sigma(t/T)$. 
The function $\sigma(r)$ is c\`{a}dl\`{a}g, non-stochastic, strictly positive, and bounded.
\end{assumption}

The principal approach to dealing with a general, slowly varying trend is to approximate the unknown trend locally by a constant.
Let $B$ be some blocklength that satisfies $2 \leq B < T$.
We divide the time series into $T-B$ overlapping blocks of length $B$ and then block-wise estimate $\rho$ via OLS under a constant trend specification.
In the fashion of \cite{schmidt1992}, as well as \cite{breitung1994}, the constant trend is estimated by the first observation in each block, which corresponds to the maximum likelihood estimator under the unit root hypothesis $\rho = 1$.
Thereafter, by pooling the $T-B$ individual block regressions, we obtain the regression equation
\begin{align*}
	\Delta y_{t+j} = \phi (y_{t+j-1} - y_j) + u_{t+j}, \qquad t=2, \ldots, B, \quad j= 1, \ldots, T-B, 
\end{align*}
where $\phi = \rho - 1$.
The pooled OLS estimator is formulated as
\begin{align*}
	\hat{\phi} = \hat{\rho} - 1 = \frac{\sum_{j=1}^{T-B} \sum_{t=2}^B \Delta y_{t+j} (y_{t+j-1} - y_j) }{\sum_{j=1}^{T-B} \sum_{t=2}^B (y_{t+j-1} - y_j )^2}. 
\end{align*}
In the following, we derive the asymptotic properties for the numerator and the denominator separately.
The numerator and denominator statistics are defined as
\begin{align*}
	\mathcal Y_{1,T} = \frac{1}{B^{3/2} T^{1/2}} \sum_{j=1}^{T-B} \sum_{t=2}^B \Delta y_{t+j} (y_{t+j-1} - y_j), \quad \mathcal Y_{2,T} = \frac{1}{B^2 T} \sum_{j=1}^{T-B} \sum_{t=2}^B (y_{t+j-1} - y_j)^2,
\end{align*}
such that $\sqrt{BT} (\hat{\rho} - 1) = \mathcal Y_{1,T}/\mathcal Y_{2,T}$.
Their counterparts without deterministics are given by
\begin{align*}
	\mathcal X_{1,T} = \frac{1}{B^{3/2} T^{1/2}} \sum_{j=1}^{T-B} \sum_{t=2}^B \Delta x_{t+j} (x_{t+j-1} - x_j), \quad \mathcal X_{2,T} = \frac{1}{B^2 T} \sum_{j=1}^{T-B} \sum_{t=2}^B (x_{t+j-1} - x_j)^2.
\end{align*}
In what follows, we show that, under the block procedure, the deterministic component can be ignored asymptotically.
All asymptotic results are jointly derived for $B,T \to \infty$.
While the statistics $\mathcal X_{1,T}$ and $\mathcal X_{2,T}$ are infeasible if $d_t$ is unknown, they can be well approximated by $\mathcal Y_{1,T}$ and $\mathcal Y_{2,T}$ in the following sense:

\begin{lemma} \label{lem_blockfilter1}
Let $\rho = 1 - c/\sqrt{BT}$ with $c \geq 0$, let $d_t$ satisfy Assumption \ref{ass:trend}, and let $u_t$ satisfy Assumption \ref{ass:errors}. Then, 
as $B,T \to \infty$, $\mathcal Y_{1,T} - \mathcal X_{1,T} = O_P(B^{-1/2})$, and $\mathcal Y_{2,T} - \mathcal X_{2,T} = O_P(T^{-1/2})$.
\end{lemma}

Accordingly, we obtain $(\mathcal Y_{1,T} - \mathcal X_{1,T}, \mathcal Y_{2,T} - \mathcal X_{2,T}) \overset{p}{\longrightarrow} (0, 0)$ jointly, and the block procedure filters out the trend component in the numerator and the denominator asymptotically.
Hence, applying Slutsky's theorem, we can write
\begin{align*}
	\sqrt{BT} (\hat{\rho} - 1) = \frac{\mathcal Y_{1,T}}{\mathcal Y_{2,T}} = \frac{\mathcal X_{1,T}}{\mathcal X_{2,T}} + o_P(1).
\end{align*}
This result is valid without any rate restrictions for $B$.
In order to obtain the limiting distribution, we formulate some properties for the numerator and denominator statistics.

\begin{lemma} \label{lem_NumDen}
Let $\rho = 1 - c/\sqrt{BT}$ with $c \geq 0$, and let $u_t$ satisfy Assumption \ref{ass:errors}. 
Then, as $B,T \to \infty$, the following statements hold true:
\begin{itemize}
\item[(a)]  $\mathcal X_{1,T} = \sum_{j=1}^T q_{j,T} - c \cdot \mathcal W_T$, where $\{q_{j,T}, \ j\leq T, \ T \in \mathbb N\}$ is a martingale difference array with $q_{j,T} = B^{-3/2} T^{-1/2} \sum_{t \in \mathcal{I}_j} \sum_{k=1}^{t-1} u_j u_{j-k}$, 
$\mathcal{I}_j = \{ t \in \mathbb{N}: \ 1 \leq t \leq B, \ j+B-T \leq t \leq j-1\}$,
and $\mathcal W_T = 0.5 \int_0^1 \sigma^2(r) \dd r + O_P(B^{1/2} T^{-1/2})$.
\item[(b)] $Var[\mathcal X_{1,T}] = \Theta(1)$ and $Var[\mathcal X_{2,T}] = \Theta(B T^{-1})$.
\item[(c)] If $c=0$ and $\sigma_t^2 = \sigma^2$ for all $t \in \mathbb N$,
\begin{align*}
	v_T^2 := \frac{\sigma^2 Var[\mathcal X_{1,T}]}{E[\mathcal X_{2,T}]} = \frac{(T-B)(2B-1)-2(B-2)}{3B(T-B)}.
\end{align*}
\end{itemize}
\end{lemma}

The previous results suggest distinguishing between different rates for $B$, which leads to two fundamentally different types of blocklength asymptotics.
The fixed-$b$ approach denotes the case where the relative blocklength $B/T$ converges to some value $b$ with $0 < b < 1$, such that $B$ and $T$ grow at the same rate.
In the small-$b$ approach, we consider a relative blocklength that converges to zero, while $B,T \to \infty$.\footnote{Note that the terminology \enquote{fixed-b and small-b asymptotics} was also used in the context of long-run variance estimation. Whereas \cite{kiefer2005} used this wording for the asymptotics of the ratio of the truncation point to the sample size, we consider the ratio of the blocklength to the sample size.}
As the blocks are overlapping, the error terms in the pooled regression equation are correlated, but, fortunately, the correlation structure is known by construction.
Together with the central limit theorem for martingale difference arrays, the following asymptotic result can be established for the small-$b$ case: 

\begin{theorem}\label{thm_asy1_sb}
Let $\rho = 1 - c/\sqrt{BT}$ with $c \geq 0$, let $d_t$ satisfy Assumption \ref{ass:trend}, and let $u_t$ satisfy Assumption \ref{ass:errors}.
Let $B/T \to 0$ as $B,T \to \infty$. 
Then, 
\begin{align*}
		\mathcal Y_{1,T}  \Dlim \mathcal{N} \bigg(-\frac{c}{2} \int_0^1 \sigma^2(r) \dd r, \ \frac{1}{3} \int_0^1 \sigma^4(r) \dd r \bigg), \quad \text{and} \ \quad \mathcal Y_{2,T}  \overset{p}{\longrightarrow} \frac{1}{2} \int_0^1 \sigma^2(r) \dd r.
\end{align*}
\end{theorem}

\noindent
Since $\mathcal Y_{2,T}$ converges in probability to a constant, we have joint convergence of $(\mathcal Y_{1,T}, \mathcal Y_{2,T})$, and the pooled estimator is asymptotically normally distributed under small-$b$ asymptotics.
Under the unit root hypothesis $\rho=1$, or, equivalently, if $c = 0$, it follows that
\begin{align*}
	\sqrt{BT}  (\hat{\rho} - 1)  \Dlim \mathcal{N} \bigg(0, \ \frac{4}{3} \frac{\int_0^1 \sigma^4(r) \dd r}{(\int_0^1 \sigma^2(r) \dd r)^2} \bigg).
\end{align*}

The asymptotic variance of $\hat \rho$ involves integrals of the second- and fourth-order powers of the function $\sigma(r)$, where the factor $\int_0^1 \sigma^4(r) \dd r/(\int_0^1 \sigma^2(r) \dd r)^2$ is equal to unity in case of homoskedasticity.
This factor also appears in the asymptotic variance matrix of the OLS estimator of the autoregressive coefficient under unconditional heteroskedasticity (see \citealt{phillips2006}).

\cite{cavaliere2005} showed that permanent changes in volatility induce a time-shift in the right-hand-side process of the functional central limit theorem.
A variance-transformed Brownian process $W_\eta(r)$ appears in the limiting distributions of Dickey-Fuller-type unit root tests.
Given the variance profile $\eta$, where $\eta(s) = ( \int_0^1 \sigma^2(r) dr )^{-1} \int_0^s \sigma^2(r) dr$, the transformed process is defined as $W_\eta(r) = W(\eta(r))$, where $W(r)$ is a standard Brownian motion.
When imposing fixed-$b$ asymptotics, the numerator and denominator statistics can be represented as a partial sum process of the innovations, which leads to the following limiting result:

\begin{theorem} \label{thm_asy1_fb}
Let $\rho = 1 - c/\sqrt{BT}$ with $c \geq 0$, let $d_t$ satisfy Assumption \ref{ass:trend}, and let $u_t$ satisfy Assumption \ref{ass:errors}.
Let $0 < b < 1$, and let $B/T \to b$ as $B,T \to \infty$. Then,
	\begin{align*}
	\lvec \mathcal Y_{1,T} \\ \mathcal Y_{2,T} \rvec 
	\Dlim 
	\lvec  0.5 b^{-3/2} \int_0^1 \sigma^2(r)  \dd r  \big(\int_0^{1-b} (J_{c,b,\eta}(b+r) - J_{c,b,\eta}(r))^2 - b(1-b) \big) \\ 
	b^{-2} \int_0^1 \sigma^2(r)  \dd r \int_0^{1-b} \int_{r}^{b+r} (J_{c,b,\eta}(s) - J_{c,b,\eta}(r))^2 \dd s \dd r \rvec,
\end{align*}
where $J_{c,b,\eta}(r) = \int_0^r e^{-(r-s)c/b} dW_\eta(s)$.
\end{theorem}

The limiting distributions are represented as functionals of the process $J_{c,b,\eta}$, which is an Ornstein-Uhlenbeck type process that is driven by a variance-transformed Wiener process.
Consequently, the pooled estimator is asymptotically represented as a functional of a standard Brownian motion.
If $\rho = 1$, the continuous mapping theorem and Theorem \ref{thm_asy1_fb} imply that
\begin{align*}
	\sqrt{BT}  (\hat{\rho} - 1) \Dlim \frac{ b^{1/2} \int_0^{1-b} \left( W_\eta(b+r) - W_\eta(r) \right)^2 \dd r + b^{3/2}(1-b)}{2 \int_0^{1-b} \int_r^{b+r} \left( W_\eta(s) - W_\eta(r) \right)^2 \dd s \dd r}
\end{align*}
under fixed-$b$ asymptotics. 
In comparison to the limiting distribution of the $\rho$-statistic in the Dickey-Fuller framework, the functional includes an additional integral, which results from pooling the block regressions.

In order to estimate the unknown parameters in the limiting distributions, we consider the residuals $\hat{u}_t = y_t - \hat{\rho} y_{t-1}$ for $t=2, \ldots, T$ and their sample mean $\ol{\hat u} = (T-1)^{-1} \sum_{j=2}^T \hat u_j$.
Let, for notational convenience, $\hat u_1 = 0$, and let
\begin{align*}
	&\hat \sigma^2 = \frac{1}{T-2} \sum_{j=2}^{T} ( \hat{u}_j - \ol{\hat u}  )^2, \quad
	\hat \kappa^2 = \frac{\sum_{j=1}^{T-B} \sum_{t=1}^B \left( \hat{u}_{j+1} - \ol{\hat u} \right)^2 \left(\hat{u}_{j+t} -  \frac{1}{B} \sum_{k=1}^B \hat{u}_{j+k} \right)^2 }{\sum_{j=1}^{T-B} \sum_{t=1}^B \Big(\hat{u}_{j+t} - \frac{1}{B} \sum_{k=1}^B \hat{u}_{j+k} \Big)^2}, \\
	 &\hat \eta(s) = \frac{  \sum_{j=2}^{\sT} \left( \hat{u}_j - \frac{1}{\sT - 1} \sum_{k=2}^{\sT} \hat{u}_{k}  \right)^2 + (sT - \sT) \left( \hat{u}_{\sT + 1} - \frac{1}{\sT } \sum_{k=2}^{\sT +1} \hat{u}_{k}  \right)^2   }{  \sum_{j=2}^{T} (\hat{u}_j - \ol{\hat u} )^2 },
\end{align*}
where $s \in [0,1]$.
We obtain the following consistency results:
\begin{lemma} \label{lem_errorvariance}
Let $\rho = 1-c/\sqrt{BT}$ with $c \geq 0$, let $d_t$ satisfy Assumption \ref{ass:trend}, and let $u_t$ satisfy Assumption \ref{ass:errors}.
\begin{itemize}
	\item[(a)] $\hat{\sigma}^2  \overset{p}{\longrightarrow} \int_0^1 \sigma^2(r) \dd r$, as $B,T \to \infty$.
	\item[(b)] $\sup_{s \in [0,1]} |\hat \eta(s) - \eta(s)| \overset{p}{\longrightarrow} 0$, as $B,T \to \infty$.
	\item[(c)] $\hat \kappa^2 \overset{p}{\longrightarrow}  \int_0^1 \sigma^4(r) \dd r / \int_0^1 \sigma^2(r) \dd r$, as $B,T \to \infty$ and $B/T \to 0$. 
\end{itemize}
\end{lemma}

\section{Pseudo t-statistics for unit root testing} \label{sec:testing} \label{sec_hettesting}

The principal concept of Dickey-Fuller-type unit root tests is to consider a $t$-test for the null hypothesis $H_0: \rho = 1$.
Following this approach in the pooled regression framework, the usual standard error is given by
$s_{\hat \rho} = \hat{\sigma} (  \sum_{j=1}^{T-B} \sum_{t=2}^B (y_{t+j-1} - y_j)^2 )^{-1/2} = \hat{\sigma} ( \mathcal Y_{2,T} B^2T )^{-1/2}$
and the conventional $t$-statistic is represented as $(\hat \rho - 1)/s_{\hat \rho} = \sqrt{B} \mathcal Y_{1,T}/\sqrt{\hat \sigma^2 \mathcal Y_{2,T}}$, which diverges in probability under $H_0$.
Accordingly, we consider a scaled pseudo $t$-statistic of the form
\begin{align}
	\tau = \frac{\hat \rho - 1}{s_{\hat \rho} \sqrt{B}} = \frac{\mathcal Y_{1,T}}{\hat \sigma \sqrt{\mathcal Y_{2,T}}},	\label{eq:tstatistic}
\end{align}
which is $O_P(1)$, as $B,T \to \infty$.

In what follows, pseudo $t$-tests are defined for both small-$b$ and fixed-$b$ block asymptotics.
In order to get a nuisance-parameter-free limiting distribution under small-$b$ asymptotics, we replace $\hat \sigma$ by $\hat \kappa$ in equation \eqref{eq:tstatistic}.
The small-$b$ pseudo $t$-statistic is given as
\begin{align*}
	\tau\text{-SB} =  \frac{\mathcal Y_{1,T}}{\hat \kappa v_T  \sqrt{\mathcal Y_{2,T}}} = \frac{\sum_{j=1}^{T-B} \sum_{t=2}^B \Delta y_{t+j} (y_{t+j-1} - y_j) }{ \hat \kappa v_T \sqrt{ B \sum_{j=1}^{T-B} \sum_{t=2}^B (y_{t+j-1} - y_j )^2 } }.
\end{align*}

The factor $v_T$ is defined in Lemma \ref{lem_NumDen}. 
Since $v_T \to 2/3$, this term provides a finite-sample correction and scales the asymptotic variance of the $t$-statistic to unity.
Under fixed-$b$ asymptotics, a nuisance term appears in the Gaussian process itself. 
By means of transforming the data with its inverse variance profile, \cite{cavaliere2007} showed that the time-transformation in the Gaussian limiting processes can be inverted.
The variance profile estimator $\hat{\eta}(s)$ is strictly increasing and admits the unique inverse function $\hat{\eta}^{-1}(s)$.
Accordingly, we consider the time-transformed series $\tilde{y}_t = y_{\lfloor \hat{\eta}^{-1}(t/T) T \rfloor}$ for $t=1, \ldots, T$.
We replace the original series in the test statistic by $\tilde y_t$ and define
\begin{align*}
	\widetilde{\mathcal Y}_{1,T}  = \frac{1}{B^{3/2} T^{1/2}} \sum_{j=1}^{T-B} \sum_{t=2}^B \Delta \tilde y_{t+j} (\tilde y_{t+j-1} - \tilde y_j), \quad 
	\widetilde{\mathcal Y}_{2,T} = \frac{1}{B^2 T} \sum_{j=1}^{T-B} \sum_{t=2}^B (\tilde y_{t+j-1} - \tilde y_j)^2,
\end{align*}
which yields the fixed-$b$ statistic 
\begin{align*}
	\tau\text{-FB} = \frac{\widetilde{\mathcal Y}_{1,T}}{\hat{\sigma} \sqrt{\widetilde{\mathcal Y}_{2,T}}}
	 = \frac{\sum_{j=1}^{T-B} \sum_{t=2}^B \Delta \tilde y_{t+j} ( \tilde y_{t+j-1} - \tilde y_j) }{\hat{\sigma} \sqrt{ B \sum_{j=1}^{T-B} \sum_{t=2}^B (\tilde y_{t+j-1} - \tilde y_j )^2 } }.
\end{align*}

In practice, the time time-transformed series $\tilde{y}_t$ can have duplicate entries in low volatility periods and therefore may not include all information of the original series in high volatility periods.
However, we do not need to discard any observations when transforming the data.
We may artificially extend the series. 
An auxiliary sample size $\widetilde T \geq T$ can be chosen in such a way that $\hat{\eta}^{-1}(t/\widetilde{T}) - \hat{\eta}^{-1}((t-1)/\widetilde{T}) \geq \widetilde T^{-1}$ for all $t = 1, \ldots, \widetilde T$.
Then, the grid of width $1/\widetilde T$ is dense enough such that $\tilde{y}_t = y_{\lfloor \hat{\eta}^{-1}(t/\widetilde{T}) \widetilde{T} \rfloor}$, $t=1, \ldots, \widetilde{T}$, includes all sample points of the original series, and the fixed-$b$ statistic may be applied to this auxiliary series.
Note that the auxiliary time series is not necessary from a theoretical point of view, but it leads to better test results in small samples.

\begin{theorem} \label{thm_tstatistics}
Let $\rho = 1-c/\sqrt{BT}$ with $c \geq 0$, let $d_t$ satisfy Assumption \ref{ass:trend}, and let $u_t$ satisfy Assumption \ref{ass:errors}.
\begin{itemize}
	\item[(a)] Let $B/T \to 0$ as $B,T \to \infty$. Then,
 \begin{align*}
 	\tau\text{-SB} \Dlim \mathcal{N} \bigg( - \frac{c  \sqrt 3}{2}  \frac{\int_0^1 \sigma^2(r) \dd r }{\sqrt{ \int_0^1 \sigma^4(r) \dd r }}, \ 1 \bigg).
\end{align*} 	
	
	\item[(b)] Let $0 < b < 1$, and let $B/T \to b$ as $B,T \to \infty$. Then,
	\begin{align*}
		\tau\text{-FB} \Dlim \frac{ \int_0^{1-b} \left( J_{c,b}(b+r) - J_{c,b}(r) \right)^2 \dd r - b(1-b)}{2 \sqrt{ b \int_0^{1-b} \int_r^{b+r} \left( J_{c,b}(s) - J_{c,b}(r) \right)^2 \dd s \dd r} },
	\end{align*}
	where $J_{c,b}(r) = \int_0^r e^{-(r-s)c/b} dW(s)$ is a standard Ornstein-Uhlenbeck process.
\end{itemize}
\end{theorem}

The unit root hypothesis is rejected in favor of stationarity if the test statistic is smaller than the $\alpha$-quantile of the limiting distribution for the case $c=0$, where $\alpha$ is the significance level.
For $\tau$-SB we can rely on standard normal quantiles as critical values.
The limiting distribution of $\tau$-FB is nonstandard.
Note that $J_c(r) = W(r)$ if $c = 0$.
Table \ref{table_critFB} presents simulated left-tailed quantiles of the null distribution for various relative blocklengths $B/T$ and significance levels.

\begin{table}[t]
\caption{Asymptotic critical values for the fixed-$b$ test}
\footnotesize
\centering
\begin{tabular}{l|lllllllll}
\multicolumn{1}{c|}{$\alpha$} & \multicolumn{9}{c}{$B/T$} \\ \hline
 & $\phantom{-}0.1$ & $\phantom{-}0.2$ & $\phantom{-}0.3$ & $\phantom{-}0.4$ & $\phantom{-}0.5$ & $\phantom{-}0.6$ & $\phantom{-}0.7$ & $\phantom{-}0.8$ & $\phantom{-}0.9$ \\ \hline
0.2 & -0.788 & -0.812 & -0.815 & -0.799 & -0.761 & -0.701 & -0.623 & -0.520 & -0.377 \\ 
0.1 & -1.126 & -1.128 & -1.104 & -1.055 & -0.987 & -0.903 & -0.798 & -0.664 & -0.486 \\ 
0.05 & -1.403 & -1.375 & -1.327 & -1.257 & -1.169 & -1.067 & -0.939 & -0.781 & -0.573 \\ 
0.04 & -1.486 & -1.446 & -1.391 & -1.318 & -1.222 & -1.113 & -0.978 & -0.814 & -0.600 \\ 
0.03 & -1.582 & -1.534 & -1.471 & -1.394 & -1.291 & -1.169 & -1.025 & -0.855 & -0.630 \\ 
0.02 & -1.709 & -1.650 & -1.579 & -1.489 & -1.374 & -1.246 & -1.094 & -0.909 & -0.669 \\ 
0.01 & -1.904 & -1.830 & -1.745 & -1.639 & -1.511 & -1.361 & -1.191 & -0.995 & -0.729 \\ 
0.001 & -2.431 & -2.320 & -2.203 & -2.042 & -1.882 & -1.692 & -1.480 & -1.226 & -0.905 \\ 
\hline
\end{tabular}
\label{table_critFB}
\parbox{12.6cm}{ \vspace{0.5ex} 
\tiny
Note: The sample paths of the standard Brownian motions contained in the asymptotic null distribution of $\tau\text{-FB}$ are simulated by a discretized version of $W(r)$ on a grid of 50,000 equidistant points. The empirical quantiles are obtained from 100,000 Monte Carlo repetitions.
}
\end{table}

From the point of view of a practitioner, the $\tau$-SB test has a number of advantages:
the distribution is standard normal; thus, there is no need to resort to new tables, and p-values are easy to implement.
In fact, the simulations in Section \ref{sec_simulations} indicate that the standard normal approximation is quite accurate in small samples if $B = \Theta(T^\gamma)$, where $0.5 \leq \gamma \leq 0.8$.
Furthermore, the unit root test is robust to heteroskedasticity without using any data modification method such as those in \cite{cavaliere2007} and \cite{beare2018} or wild bootstrap implementations (see \citealt{cavaliere2008bootstrap}).

\section{Testing under short-run dynamics} \label{sec_serialcorrelation}

A more realistic scenario for macroeconomic variables is that error terms are serially correlated.
We impose the following assumption on the error process:

\begin{assumption}[serially correlated errors] \label{ass_lrv}
The process $\{u_t\}_{t \in \mathbb{Z}}$ possesses the moving average representation $u_t = \psi(L) \epsilon_t = \sum_{i=0}^\infty \psi_i \epsilon_{t-i}$ with $\sum_{i=0}^\infty |\psi_i| < \infty$, where $L$ is the usual lag operator.
Moreover, all solutions $z$ of the equation $\psi(z) = 0$ satisfy $|z| > 1$.
The process $\{\epsilon_t\}_{t \in \mathbb{Z}}$ is independently distributed with $E[\epsilon_t]= 0$, $E[\epsilon_t^2] = \sigma_t^2$ and $E[\epsilon_t^4] < \infty$, where $\sigma_t = \sigma(t/T)$. 
The function $\sigma(r)$ is c\`{a}dl\`{a}g, non-stochastic, strictly positive, and bounded.
\end{assumption}

\noindent
Assumption \ref{ass_lrv} implies that the moving average representation of $u_t$ is invertible, and we may write $\theta(L) u_t = u_t - \sum_{i=1}^\infty \theta_i u_{t-i} = \epsilon_t$,
where $\theta(z) = 1 - \sum_{i=1}^\infty \theta_i z^i$, and $\sum_{i=1}^\infty |\theta_i| < \infty$.
In order to correct for the effect of short-run dynamics, we follow \cite{breitung2005}, among others, and consider the pre-whitened series $x_t^* = \theta(L) x_t$.
By equation \eqref{eq:model}, it follows that
\begin{align*}
	x_t^* = \theta(L) \rho x_{t-1} + \theta(L) u_t = \rho x_{t-1}^* + \epsilon_t,
\end{align*}
where $\epsilon_t$ satisfies the same conditions as $u_t$ under Assumption \ref{ass:errors}.
Consequently, if the unit root statistics are defined in terms of
\begin{align*}
	\mathcal X_{1,T}^* = \frac{1}{B^{3/2} T^{1/2}} \sum_{j=1}^{T-B} \sum_{t=2}^B \Delta x^*_{t+j} (x^*_{t+j-1} - x^*_j), \quad 
	\mathcal X_{2,T}^* = \frac{1}{B^2 T} \sum_{j=1}^{T-B} \sum_{t=2}^B (x^*_{t+j-1} - x^*_j)^2
\end{align*}
instead of $\mathcal X_{1,T}$ and $\mathcal X_{2,T}$, their limiting distributions coincide with those presented in the previous sections.

Since the autoregressive parameters of the error process are unknown, they need to be estimated.
In the fashion of \cite{said1984} and \cite{chang2002}, we fix some lag order $p_T$ and consider the AR($p_T$) error representation
$u_t = \sum_{i=1}^{p_T} \theta_i u_{t-i} + \epsilon_{p_T,t}$ with $\epsilon_{p_T,t} = \sum_{i=p_T+1}^\infty \theta_i u_{t-i} + \epsilon_t$.
Then,
\begin{align}
	\Delta x_t = \phi x_{t-1} + \sum_{i=1}^{p_T} \theta_i u_{t-i} + \epsilon_{p_T,T}, \label{eq:augmentedregression}
\end{align}
which is equal to $\sum_{i=1}^{p_T} \theta_i \Delta x_{t-i} + \epsilon_{p_T,T}$ under the unit root hypothesis.
The lag order $p_T$ is allowed to grow with the sample size $T$.
In what follows, we show that the differenced deterministic terms are asymptotically negligible, as $p_T \to \infty$ with $p_T=o(B^{1/2})$, and we may replace $\Delta x_{t-i}$ by $\Delta y_{t-i}$ for all $i \geq 0$ in the augmented regression equation.
Let $(\hat \varphi$, $\hat \theta_1, \ldots, \hat \theta_{p_T})'$ be the least squares coefficient vector from the regression of $\Delta y_t$ on $y_{t-1}, \Delta y_{t-1}, \ldots, \Delta y_{t-{p_T}}$, for $t = p_T+1. \ldots, T$.

\begin{lemma} \label{lem_prewhitenconsistency}
Let $\rho = 1 - c/\sqrt{BT}$ with $c \geq 0$, let $d_t$ satisfy Assumption \ref{ass:trend}, and let $u_t$ satisfy Assumption \ref{ass_lrv}. 
Then, $\sum_{i=1}^{p_T} (\hat \theta_i - \theta_i) = O_P(p_T B^{-1/2})$, as $p_T,B,T \to \infty$.
\end{lemma}

The estimated pre-whitened series is defined as $\hat y_{t}^* = y_t - \sum_{i=1}^{p_T} \hat \theta_i y_{t-i}$,
and the corresponding numerator and denominator statistics are given by 
\begin{align*}
	\hat{\mathcal Y}_{1,T}^*  = \frac{1}{B^{3/2} T^{1/2}} \sum_{j=1}^{T-B} \sum_{t=2}^B \Delta \hat y^*_{t+j} (\hat y^*_{t+j-1} - \hat y^*_j), \quad 
	\hat{\mathcal Y}_{2,T}^* = \frac{1}{B^2 T} \sum_{j=1}^{T-B} \sum_{t=2}^B (\hat y^*_{t+j-1} - \hat y^*_j)^2.
\end{align*}

\begin{lemma} \label{lem_blockfilter4}
Let $\rho = 1 - c/\sqrt{BT}$ with $c \geq 0$, let $d_t$ satisfy Assumption \ref{ass:trend}, and let $u_t$ satisfy Assumption \ref{ass_lrv}.
Then, $\hat{\mathcal Y}_{1,T}^* - \mathcal X_{1,T}^* = O_P(p_T B^{-1/2})$, and $\hat{\mathcal Y}_{2,T}^* - \mathcal X_{2,T}^* = O_P(p_T T^{-1/2})$, as $p_T, B,T \to \infty$.
\end{lemma}

As a direct consequence, $(\hat{\mathcal Y}_{1,T}^* - \mathcal X_{1,T}^*, \hat{\mathcal Y}_{2,T}^* - \mathcal X_{2,T}^*) \overset{p}{\longrightarrow} (0, 0)$ if $p_T=o(B^{1/2})$.
Let $\hat \rho^*$ be given by $\sqrt{BT}(\hat \rho^* - 1) = \hat{\mathcal Y}_{1,T}^*/\hat{\mathcal Y}_{2,T}^*$ and let the pre-whitened residuals be defined as $\hat u_t^* = \hat y_t^* - \hat \rho^* \hat y_{t-1}^*$, for $t = p_T+1, \ldots, T$.
For notational convenience, let $\hat u_1^* = \ldots = \hat u_{p_T}^* = 0$.
The pre-whitened counterparts of the estimators from Lemma \ref{lem_errorvariance} are defined as
\begin{align*}
	&\hat \sigma^{*2} = \frac{1}{T-2} \sum_{j=2}^{T} ( \hat{u}_j^* - \ol{\hat u^*}  )^2, \quad
	\hat \kappa^{*2} = \frac{\sum_{j=1}^{T-B} \sum_{t=1}^B \left( \hat{u}_{j+1}^* - \ol{\hat u^*} \right)^2 \left(\hat{u}_{j+t}^* -  \frac{1}{B} \sum_{k=1}^B \hat{u}_{j+k}^* \right)^2 }{\sum_{j=1}^{T-B} \sum_{t=1}^B \Big(\hat{u}_{j+t}^* - \frac{1}{B} \sum_{k=1}^B \hat{u}_{j+k}^* \Big)^2}, \\
	 &\hat \eta^*(s) = \frac{  \sum_{j=2}^{\sT} \left( \hat{u}_j^* - \frac{1}{\sT - 1} \sum_{k=2}^{\sT} \hat{u}_{k}^*  \right)^2 + (sT - \sT) \left( \hat{u}_{\sT + 1}^* - \frac{1}{\sT } \sum_{k=2}^{\sT +1} \hat{u}_{k}^*  \right)^2   }{  \sum_{j=2}^{T} (\hat{u}_j^* - \ol{\hat u^*} )^2 }.
\end{align*}
Analogously, we consider the time-transformed pre-whitened series $\tilde{y}_t^* = \hat y_{\lfloor \hat{\eta}^{*-1}(t/T) T \rfloor}^*$ for all $t=1, \ldots, T$, 
where $\hat{\eta}^{*-1}(s)$ is the unique inverse of $\hat \eta^*(s)$,
and we define 
\begin{align*}
	\widetilde{\mathcal Y}_{1,T}^*  = \frac{1}{B^{3/2} T^{1/2}} \sum_{j=1}^{T-B} \sum_{t=2}^B \Delta \tilde y_{t+j}^* (\tilde y_{t+j-1}^* - \tilde y_j^*), \quad 
	\widetilde{\mathcal Y}_{2,T}^* = \frac{1}{B^2 T} \sum_{j=1}^{T-B} \sum_{t=2}^B (\tilde y_{t+j-1}^* - \tilde y_j^*)^2.
\end{align*}
For any lag order $p_T\geq 0$, the pre-whitened versions of the test statistics are given by
\begin{align*}
	\tau\text{-SB}_{p_T} = \frac{\hat{\mathcal Y}_{1,T}^*}{\hat \kappa^* v_T \sqrt{\hat{\mathcal Y}_{2,T}^*}}, \quad
	\tau\text{-FB}_{p_T} =  \frac{\hat{\mathcal Y}_{1,T}^*}{\hat{\sigma}^* \sqrt{\hat{\mathcal Y}_{2,T}^*}}.
\end{align*}
Note that $\tau\text{-SB}_0 = \tau\text{-SB}$ and $\tau\text{-FB}_0 = \tau\text{-FB}$.
To summarize, we obtain the following limiting distributions:

\begin{theorem} \label{thm_shortrun}
Let $\rho = 1- c/\sqrt{BT}$, let $d_t$ satisfy Assumption \ref{ass:trend}, and let $u_t$ satisfy Assumption \ref{ass_lrv}. Furthermore, let $p_T = o(B^{1/2})$.
\begin{itemize}
	\item[(a)] Let $B/T \to 0$ as $B,T \to \infty$. Then, $\hat \kappa^{*2} \overset{p}{\longrightarrow} \int_0^1 \sigma^4(r) \dd r / \int_0^1 \sigma^2(r) \dd r$, and 
	\begin{align*}
		\tau\text{-SB}_{p_T} \Dlim \mathcal{N} \bigg( - \frac{c  \sqrt 3}{2}  \frac{\int_0^1 \sigma^2(r) \dd r }{\sqrt{ \int_0^1 \sigma^4(r) \dd r }}, \ 1 \bigg).
	\end{align*}		
	\item[(b)] Let $0 < b < 1$, and let $B/T \to b$ as $B,T \to \infty$. Then, $\sup_{r \in [0,1]} |\hat \eta(s) - \eta(s)| \overset{p}{\longrightarrow} 0$, $\hat \sigma^{*2} \overset{p}{\longrightarrow} \int_0^1 \sigma^2(r) \dd r$, and
	\begin{align*}
		\tau\text{-FB}_{p_T} \Dlim \frac{ \int_0^{1-b} \left( J_{c,b}(b+r) - J_{c,b}(r) \right)^2 \dd r - b(1-b)}{2 \sqrt{ b \int_0^{1-b} \int_r^{b+r} \left( J_{c,b}(s) - J_{c,b}(r) \right)^2 \dd s \dd r} },
	\end{align*}
	where $J_{c,b}(r) = \int_0^r e^{-(r-s)c/b} dW(s)$.
\end{itemize}	
\end{theorem}

The lag order $p_T$ is typically unknown in practice and can be chosen using conventional lag order selection methods, such as the Bayesian information criterion (BIC) or by the general-to-specific methodology in the fashion of \cite{ng1995}. 
The maximum lag order $p_{max}$ can be chosen for instance by the rule of thumb provided by \cite{schwert1989}.
For the special case of a single break in the deterministic component, \cite{demetrescu2016} showed that if $p_T$ is determined by a usual information criterion the correct lag length is selected asymptotically.

\section{Simulations} \label{sec_simulations}

In this section, the finite sample performance of the unit root tests is evaluated by means of Monte Carlo simulations.
The analysis includes different specifications for both the deterministic part $d_t$ and the stochastic part $x_t$.

\begin{table}[t] 
\caption{Trend functions}
\footnotesize
\centering
\begin{tabular}{c|l|l}  
 & type of the trend & functional form \\
\hline
1 & sharp break & $d(r) = \lambda \cdot 1_{\{ r\leq 2/3 \}}$ \\
2 & u-shaped break & $d(r) = \lambda \cdot 1_{\{ r\leq 1/4 \}} + \lambda \cdot 1_{\{ r > 3/4 \}}$ \\
3 & continuous break & $d(r) = \lambda \cdot  (4 r \cdot 1_{\{ r > 2/3 \}} - 8/3)$ \\
4 & u-shaped break in intercept & $d(r) = \lambda \cdot ( r 1_{\{ r\leq 1/4 \}} + (r-1) 1_{\{ 1/4 < r \leq 3/4 \}} + r 1_{\{ t > 3/4 \}} )$ \\
5 & LSTAR break & $d(r) = \lambda \cdot (1 + \exp(20(r-0.75)))^{-1} $ \\
6 & offsetting LSTAR break & $d(r) = \lambda /(1 + \exp(20(r-0.2))) - 0.5 \lambda/(1 + \exp(20(r-0.75))) $  \\ 
7 & triangular break & $d(r) = \lambda \cdot ( 2r 1_{\{ r\leq 1/2 \}} + 2(1-r) 1_{\{ r > 1/2 \}} )$ \\
8 & Fourier break & $d(r) = \lambda \cdot  0.5 \cos(2 \pi r)  $ \\
\hline
\end{tabular}
\label{tab_trends}
\parbox{14.2cm}{ \vspace{0.5ex} 
\tiny
Note: The functional form of the trend functions for the simulations are presented. The parameter $\lambda$ determines the size of the trend.
}
\end{table}

\begin{figure}[t]
\caption{Plots of the trend functions}
\vspace{-2ex}
\begin{center}
\includegraphics[scale=0.55]{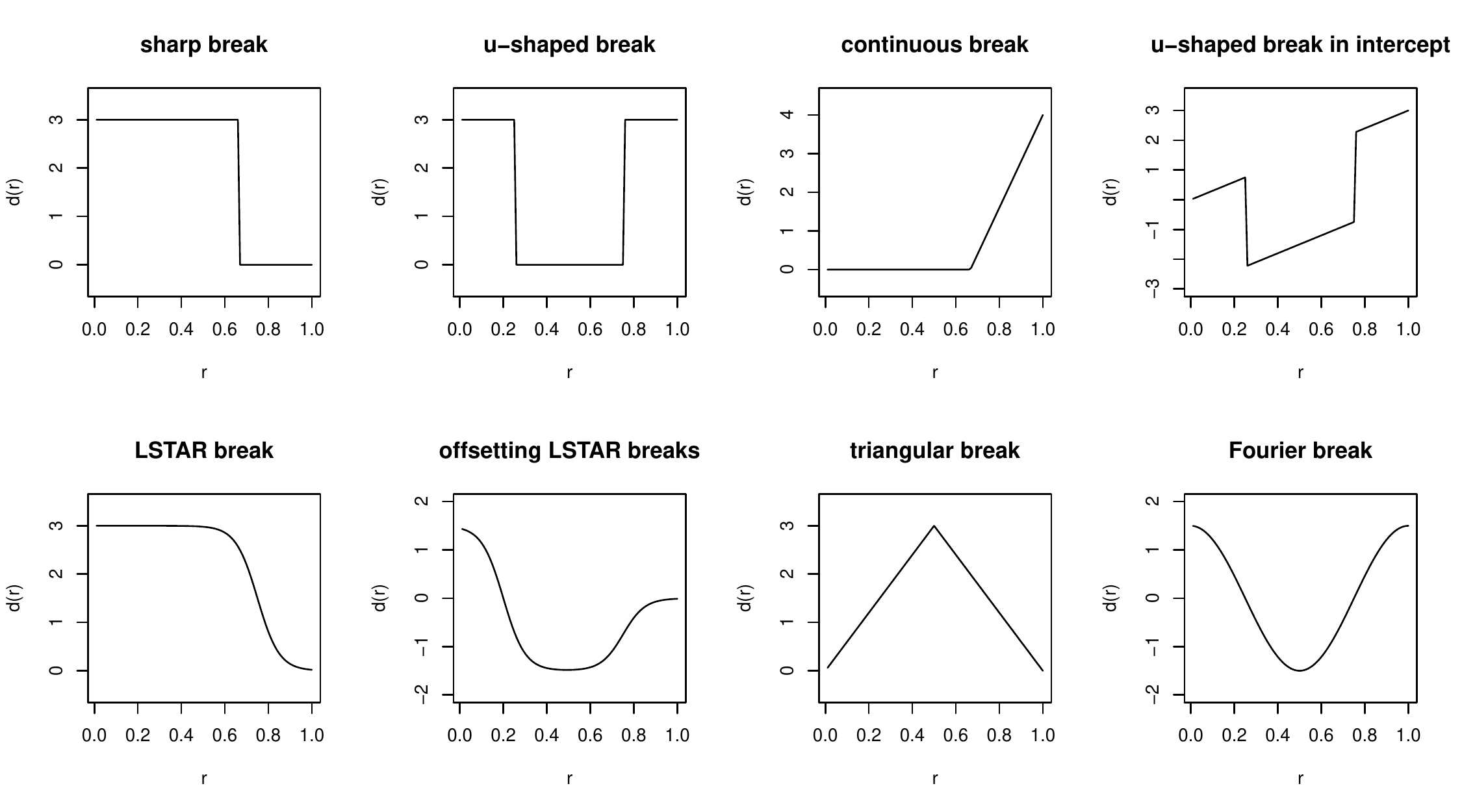}
\end{center}
\label{fig_trends}
\parbox{12.5cm}{ \vspace{-6ex} 
\tiny
Note: The plots of the of the trend functions from Table \ref{tab_trends} are presented. The trend size is $\lambda = 3$.
}
\end{figure}

While the zero-trend $d_t = 0$ is the main benchmark, we consider several other trends including sharp breaks and smooth changes of different shapes. 
The trend specifications are presented in Table \ref{tab_trends} and Figure \ref{fig_trends}. 
The parameter $\lambda$ determines the size of the break. 
Similar trend functions are also considered in \cite{jones2014} in order to evaluate the performance of the unit root test by \cite{enders2012a}.

The stochastic part $x_t$ is simulated both under the null hypothesis $\rho = 1$ and the alternative hypothesis $\rho = 0.9$.
For the errors $u_t$, we consider an independent process as well as the AR(1) process $u_t = 0.5 u_{t-1} + \epsilon_t$ with standard normal innovations.
Furthermore, results with heteroskedastic innovations using the variance function $\sigma^2(r) = 1+\lambda \cdot 1_{\{ r\leq 2/3 \}}$ are presented.

The small-$b$ tests are implemented using blocklengths of the form $B=T^\gamma$ with parameters $\gamma \in \{ 0.5, 0.6, 0.7, 0.8 \}$.
For the fixed-$b$ versions, we consider $B = b \cdot T$ with relative blocklengths $b \in \{0.2, 0.4, 0.6\}$.
For all tests, the lag augmentation order $p_T$ is either fixed or flexibly determined by the BIC with a maximum lag order of $p_{max} = 5$.
All empirical size levels are presented for a significance level of 5\%, and the models are simulated with 100,000 repetitions for sample sizes of $T=100$ and $T=300$.
As noted by \cite{muller2003}, the power of a unit root test depends on the initial condition, and the initial value is simulated as $x_0 \sim \mathcal{N}(0,\sigma_0^2)$ for $\sigma_0^2 \in \{0, 5, 10\}$.

In order to demonstrate the advantage of the fixed-$b$ and small-$b$ unit root tests, their finite sample results are compared to those obtained by conventional unit root tests.
As the main benchmark, we consider the augmented Dickey-Fuller test by \cite{said1984} with constant trend specification (ADF henceforth),
which is the $t$-test for the hypothesis $\phi = 0$ in the regression $\Delta y_t = \phi y_{t-1} + \beta_0 + \sum_{i=1}^{p_T} \xi_i \Delta y_{t-i} + e_t$.

\cite{elliott1996} proposed a feasible point-optimal test with local-to-unity GLS demeaning in the ADF regression.
Let the deterministic trend function be given by the vector $z_t$, and let $\alpha^* = 1 - \ol c/T$, where $\ol c \in \mathbb{R}$.
Furthermore, let $y_{\ol c, t} = y_t - \alpha^* y_{t-1}$ and $Z_{\ol c, t} = z_t - \alpha^* z_{t-1}$ for $t \geq 2$, and let $y_{\ol c, 1} = y_1$ and $Z_{\ol c, 1} = z_1$.
The Dickey-Fuller GLS test is then the $t$-test for the hypothesis $\phi = 0$ in the regression $\Delta y_t^d = \phi y_{t-1}^d + \sum_{i=1}^{p_T} \xi_i \Delta y_{t-i}^d + e_t$,
where $y_t^d = y_t - \hat \beta' z_t$ and where $\hat \beta$ is the OLS estimator from a regression of $y_{\ol c, t}$ on $Z_{\ol c, t}$.
For the constant trend specification (DF-GLS henceforth), we set $z_t = 1$ and $\ol c = 7$, and, for the linear trend specification (DF-GLS-trend henceforth), $z_t = (1,t)'$ and $\ol c = 13.5$ are considered. 
Note that the point-optimal test with GLS demeaning is asymptotically equivalent with the Dickey-Fuller test for $d_t = 0$ computed using the series with initial value subtraction (see \citealt{elliott1996})

An approach that does not assume a precise model for the trend component is that developed by \citet{enders2012a} (EL henceforth).
A flexible Fourier form is used to approximate smooth breaks in the trend function. Structural changes can be captured by the low frequency components of a series. 
In its simplest form, \citet{enders2012a} considered the parametric trend model
$d(r) =  \alpha_0 +  \gamma r + \alpha_1 \sin(2 \pi r) + \beta_1 \cos(2 \pi r)$.
More frequencies could be included, but doing so could lead to an over-fitting problem.
The test works as follows:
First, the auxiliary regression $\Delta y_t = \delta_0 + \delta_{1} \Delta \sin(2 \pi t/T) + \delta_{2} \Delta \cos(2 \pi t/T) + v_t$
is considered with OLS estimates $\widehat{\delta}_0$, 
$\widehat{\delta}_{1}$, and $\widehat{\delta}_{2}$. 
Let $\widetilde D_t = \hat \delta_0 t + \hat \delta_1 \sin(2 \pi t/T) + \hat \delta_2 \cos (2 \pi t / T)$, which yields the detrended series $\widetilde{S}_t = y_t - \widetilde D_t - (y_1 - \widetilde D_1)$.
Finally, the test statistic is given by the $t$-statistic for the null hypothesis $\phi = 0$ in the regression
$\Delta y_t = \phi \widetilde{S}_{t-1} + \beta_0 +  \beta_{1} \Delta \sin(2 \pi t/T) + \beta_{2} \Delta \cos(2 \pi t/T) + \sum_{i=1}^{p_T} \xi_i \Delta \widetilde{S}_{t-i} + e_t$.
	
\begin{figure}[ht]
\caption{Effect of the initial condition on the finite-sample power}
\vspace{-4ex}
\begin{center}
\includegraphics[scale=0.22]{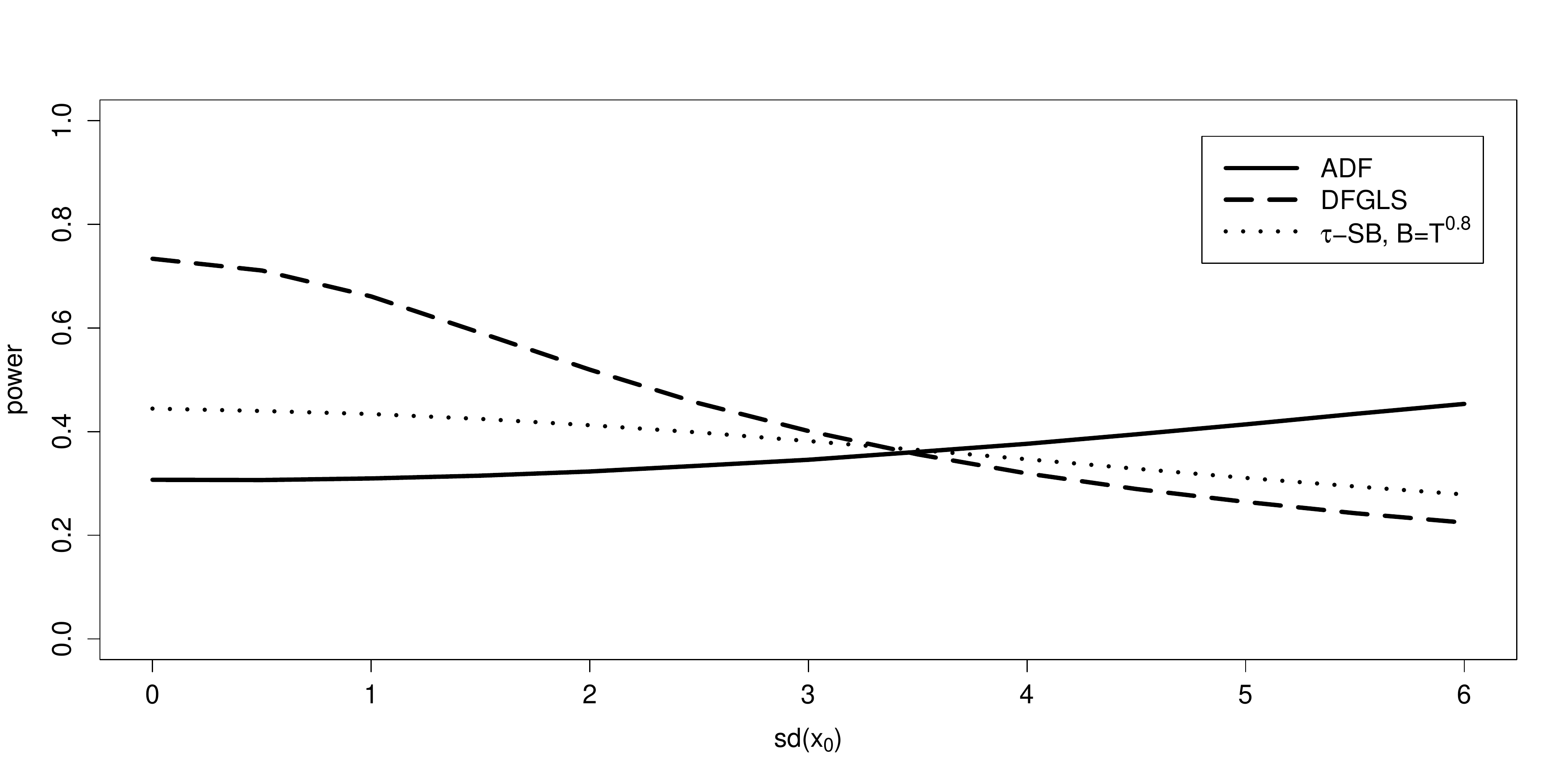}
\end{center}
\label{fig_initial}
\parbox{11.5cm}{ \vspace{-4ex} 
\tiny
Note: Size-adjusted power results for different tests are presented.
The initial condition is simulated from a normal distribution with mean zero and different values for $\sigma_0^2 = Var[x_0]$, where $\sigma_0$ is shown on the x-axis.
The simulation results are reported for for a nominal size level of $5\%$, for 100,000 replications with $T=100$, $\rho = 0.9$, the zero trend specification $d_t = 0$, and independent standard normal innovations $u_t$.
}
\end{figure}

\citet{harvey2005, harvey2006} showed that, if $x_0 \sim \mathcal{N}(0, \sigma_\alpha^2/(1-\rho^2))$ for $\rho = 1-c/T$ with $c > 0$ and some $\sigma_\alpha > 0$, the limiting distributions of the ADF and the DF-GLS test depend on the additional nuisance parameter $\sigma_\alpha$.
The DF-GLS test is optimal for the zero initial condition $x_0 = 0$, but its power decreases monotonically in $\sigma_\alpha$, while the power of the ADF test increases.
Figure \ref{fig_initial} indicates that the pooled tests are less sensitive to this effect across different values of $\sigma_\alpha$.
Furthermore, there is no test that outperforms the other tests uniformly across $\sigma_\alpha$ for this situation in terms of size-adjusted power.

\begin{table}[ht]
\caption{Size and power results under the zero-trend specification}
\centering
\begin{scriptsize}
\begin{tabular}{l|cc|cc|cc|cc|cc|cc}
\multicolumn{1}{c|}{initial value} & \multicolumn{4}{c|}{$x_0 = 0$} & \multicolumn{4}{c|}{$x_0 \sim \mathcal N(0,5)$} & \multicolumn{4}{c}{$x_0 \sim \mathcal N(0,10)$}  \\ \hline
\multicolumn{1}{c|}{sample size} &  \multicolumn{2}{c|}{$T=100$} & \multicolumn{2}{c|}{$T=300$} &  \multicolumn{2}{c|}{$T=100$} & \multicolumn{2}{c|}{$T=300$} &  \multicolumn{2}{c|}{$T=100$} & \multicolumn{2}{c}{$T=300$} \\
\multicolumn{1}{c|}{$\rho$} & $1$ & $0.9$ &  $1$ & $0.9$ &  $1$ & $0.9$  &  $1$ & $0.9$ &  $1$ & $0.9$  &  $1$ & $0.9$ \\ 
\hline
\multicolumn{13}{l}{\textbf{i.i.d.\ errors -- no lag augmentation (p=0)}} \\ \hline
$\tau$-SB, $B = T^{0.5}$   & 0.063 & 0.346 & 0.057 & 0.870 & 0.064 & 0.329 & 0.057 & 0.864 & 0.064 & 0.315 & 0.057 & 0.859 \\ 
$\tau$-SB, $B = T^{0.6}$   & 0.064 & 0.407 & 0.059 & 0.963 & 0.064 & 0.388 & 0.059 & 0.961 & 0.064 & 0.371 & 0.059 & 0.959 \\ 
$\tau$-SB, $B = T^{0.7}$    & 0.062 & 0.459 & 0.058 & 0.992 & 0.061 & 0.434 & 0.059 & 0.991 & 0.061 & 0.413 & 0.059 & 0.990 \\ 
$\tau$-SB, $B = T^{0.8}$   & 0.049 & 0.428 & 0.048 & 0.996 & 0.049 & 0.400 & 0.049 & 0.995 & 0.049 & 0.375 & 0.049 & 0.995 \\ 
$\tau$-FB, $B = 0.2T$    & 0.042 & 0.306 & 0.046 & 0.973 & 0.041 & 0.287 & 0.046 & 0.972 & 0.041 & 0.270 & 0.046 & 0.970 \\ 
$\tau$-FB, $B = 0.4T$   & 0.047 & 0.374 & 0.047 & 0.989 & 0.046 & 0.346 & 0.048 & 0.988 & 0.047 & 0.323 & 0.048 & 0.987 \\ 
$\tau$-FB, $B = 0.6T$   & 0.047 & 0.386 & 0.046 & 0.989 & 0.047 & 0.350 & 0.046 & 0.988 & 0.047 & 0.320 & 0.046 & 0.986 \\ 
ADF & 0.054 & 0.329 & 0.052 & 0.996 & 0.054 & 0.348 & 0.050 & 0.996 & 0.054 & 0.367 & 0.050 & 0.996 \\ 
DF-GLS & 0.078 & 0.792 & 0.058 & 1.000 & 0.077 & 0.617 & 0.058 & 0.947 & 0.077 & 0.516 & 0.058 & 0.858 \\ 
DF-GLS-trend & 0.069 & 0.371 & 0.053 & 0.994 & 0.069 & 0.324 & 0.052 & 0.955 & 0.069 & 0.292 & 0.052 & 0.894 \\ 
EL & 0.061 & 0.140 & 0.054 & 0.775 & 0.061 & 0.134 & 0.053 & 0.755 & 0.061 & 0.130 & 0.053 & 0.732 \\  
\hline 
\multicolumn{13}{l}{\textbf{AR(1) errors -- fixed lag augmentation (p=1)}} \\ \hline
$\tau\text{-SB}_1$, $B = T^{0.5}$   & 0.012 & 0.125 & 0.021 & 0.679 & 0.012 & 0.124 & 0.022 & 0.675 & 0.012 & 0.121 & 0.022 & 0.674 \\ 
$\tau\text{-SB}_1$, $B = T^{0.6}$   & 0.025 & 0.222 & 0.038 & 0.877 & 0.025 & 0.220 & 0.038 & 0.876 & 0.025 & 0.216 & 0.038 & 0.875 \\ 
$\tau\text{-SB}_1$, $B = T^{0.7}$    & 0.038 & 0.305 & 0.046 & 0.958 & 0.037 & 0.301 & 0.046 & 0.957 & 0.037 & 0.297 & 0.046 & 0.957 \\ 
$\tau\text{-SB}_1$, $B = T^{0.8}$   & 0.034 & 0.290 & 0.042 & 0.972 & 0.033 & 0.286 & 0.042 & 0.972 & 0.033 & 0.281 & 0.042 & 0.972 \\ 
$\tau\text{-FB}_1$, $B = 0.2T)$   & 0.025 & 0.189 & 0.040 & 0.922 & 0.025 & 0.187 & 0.040 & 0.922 & 0.025 & 0.184 & 0.040 & 0.922 \\ 
$\tau\text{-FB}_1$, $B = 0.4T$   & 0.037 & 0.270 & 0.044 & 0.960 & 0.037 & 0.268 & 0.045 & 0.961 & 0.037 & 0.263 & 0.045 & 0.960 \\ 
$\tau\text{-FB}_1$, $B = 0.6T$   & 0.039 & 0.281 & 0.044 & 0.962 & 0.038 & 0.276 & 0.044 & 0.962 & 0.037 & 0.272 & 0.044 & 0.961 \\ 
ADF & 0.056 & 0.263 & 0.051 & 0.970 & 0.056 & 0.267 & 0.051 & 0.971 & 0.056 & 0.271 & 0.051 & 0.972 \\ 
DF-GLS & 0.077 & 0.722 & 0.058 & 1.000 & 0.077 & 0.656 & 0.058 & 0.993 & 0.077 & 0.602 & 0.058 & 0.973 \\ 
DF-GLS-trend & 0.071 & 0.309 & 0.052 & 0.970 & 0.071 & 0.297 & 0.052 & 0.956 & 0.071 & 0.285 & 0.052 & 0.937 \\ 
EL & 0.067 & 0.125 & 0.056 & 0.636 & 0.068 & 0.125 & 0.056 & 0.628 & 0.068 & 0.123 & 0.056 & 0.620 \\     
\hline
\multicolumn{13}{l}{\textbf{AR(1) errors -- flexible lag augmentation (p determined by BIC)}} \\ \hline
$\tau\text{-SB}_p$, $B = T^{0.5}$   & 0.006 & 0.093 & 0.016 & 0.680 & 0.006 & 0.093 & 0.016 & 0.676 & 0.006 & 0.091 & 0.016 & 0.674 \\ 
$\tau\text{-SB}_p$, $B = T^{0.6}$   & 0.018 & 0.200 & 0.033 & 0.873 & 0.018 & 0.198 & 0.034 & 0.872 & 0.018 & 0.195 & 0.034 & 0.871 \\ 
$\tau\text{-SB}_p$, $B = T^{0.7}$    & 0.032 & 0.296 & 0.044 & 0.952 & 0.032 & 0.293 & 0.044 & 0.953 & 0.031 & 0.289 & 0.044 & 0.952 \\ 
$\tau\text{-SB}_p$, $B = T^{0.8}$   & 0.032 & 0.287 & 0.042 & 0.968 & 0.030 & 0.284 & 0.041 & 0.968 & 0.030 & 0.280 & 0.041 & 0.968 \\ 
$\tau\text{-FB}_p$, $B = 0.2T$    & 0.020 & 0.171 & 0.038 & 0.916 & 0.020 & 0.170 & 0.038 & 0.917 & 0.020 & 0.168 & 0.038 & 0.916 \\ 
$\tau\text{-FB}_p$, $B = 0.4T$   & 0.033 & 0.254 & 0.043 & 0.956 & 0.033 & 0.254 & 0.044 & 0.956 & 0.033 & 0.250 & 0.044 & 0.956 \\ 
$\tau\text{-FB}_p$, $B = 0.6T$   & 0.035 & 0.263 & 0.044 & 0.957 & 0.034 & 0.261 & 0.043 & 0.957 & 0.033 & 0.258 & 0.043 & 0.956 \\ 
ADF & 0.058 & 0.269 & 0.051 & 0.969 & 0.059 & 0.272 & 0.052 & 0.970 & 0.059 & 0.276 & 0.052 & 0.971 \\ 
DF-GLS & 0.085 & 0.703 & 0.060 & 0.999 & 0.084 & 0.637 & 0.059 & 0.991 & 0.084 & 0.584 & 0.059 & 0.967 \\ 
DF-GLS-trend & 0.082 & 0.317 & 0.054 & 0.960 & 0.081 & 0.302 & 0.055 & 0.943 & 0.081 & 0.289 & 0.055 & 0.921 \\ 
EL & 0.106 & 0.175 & 0.066 & 0.637 & 0.106 & 0.173 & 0.064 & 0.628 & 0.106 & 0.171 & 0.064 & 0.621 \\ 
\hline
\end{tabular} 
\end{scriptsize}
\parbox{16cm}{ \vspace{0.5ex} 
\tiny
Note: Simulation results are reported for 100,000 replications. The zero-trend $d_t = 0$ is considered for all $t = 1, \ldots, T$.
	The AR(1) process is given by $u_t = 0.5 u_{t-1} + \epsilon_t$.
		All innovations are simulated independently as standard normal random variables.
For the small-$b$ and fixed-$b$ tests, the lag order $p$ refers to the pre-whitening scheme, and, for the conventional tests, $p$ represents the augmentation order.
The rejection frequencies are based on the asymptotic critical values for a significance level of $5\%$.
}
\label{tab_notrend}
\end{table}

\begin{table}[hp]
\caption{Size and power results under different trends and i.i.d.\ errors (1/2)}
\centering
\begin{scriptsize}
\begin{tabular}{l|ccc|ccc|ccc|ccc}
\multicolumn{1}{c|}{sample size} &  \multicolumn{6}{c|}{$T=100$}  &  \multicolumn{6}{c}{$T=300$}  \\ \hline
\multicolumn{1}{c|}{$\rho$} &  \multicolumn{3}{c|}{$\rho = 1$} & \multicolumn{3}{c|}{$\rho = 0.9$} &  \multicolumn{3}{c|}{$\rho = 1$}  &  \multicolumn{3}{c}{$\rho = 0.9$}  \\
\multicolumn{1}{c|}{$\lambda$} & $3$ & $6$ & $9$ & $3$ & $6$ & $9$ & $3$ & $6$ & $9$ & $3$ & $6$ & $9$  \\ 
\hline 
\multicolumn{3}{l}{\textbf{sharp break}} \\ \hline 
$\tau$-SB, $B = T^{0.5}$   & 0.064 & 0.064 & 0.063 & 0.281 & 0.194 & 0.129 & 0.057 & 0.058 & 0.058 & 0.837 & 0.752 & 0.623 \\ 
$\tau$-SB, $B = T^{0.6}$   & 0.065 & 0.067 & 0.068 & 0.318 & 0.198 & 0.114 & 0.059 & 0.061 & 0.062 & 0.941 & 0.861 & 0.705 \\ 
$\tau$-SB, $B = T^{0.7}$ & 0.063 & 0.068 & 0.072 & 0.322 & 0.155 & 0.069 & 0.059 & 0.060 & 0.063 & 0.976 & 0.885 & 0.638 \\ 
$\tau$-SB, $B = T^{0.8}$   & 0.069 & 0.117 & 0.153 & 0.319 & 0.189 & 0.108 & 0.051 & 0.056 & 0.063 & 0.966 & 0.709 & 0.241 \\ 
$\tau$-FB, $B = 0.2T$    & 0.041 & 0.043 & 0.041 & 0.218 & 0.129 & 0.058 & 0.046 & 0.044 & 0.043 & 0.936 & 0.758 & 0.474 \\ 
$\tau$-FB, $B = 0.4T$   & 0.044 & 0.027 & 0.011 & 0.220 & 0.060 & 0.009 & 0.048 & 0.043 & 0.033 & 0.940 & 0.654 & 0.237 \\ 
$\tau$-FB, $B = 0.6T$   & 0.042 & 0.022 & 0.006 & 0.225 & 0.055 & 0.004 & 0.046 & 0.040 & 0.027 & 0.936 & 0.639 & 0.205 \\ 
ADF & 0.050 & 0.038 & 0.023 & 0.169 & 0.021 & 0.001 & 0.049 & 0.045 & 0.038 & 0.898 & 0.247 & 0.004 \\ 
DF-GLS & 0.078 & 0.075 & 0.065 & 0.402 & 0.105 & 0.011 & 0.059 & 0.059 & 0.059 & 0.885 & 0.599 & 0.142 \\ 
DF-GLS-trend & 0.069 & 0.067 & 0.055 & 0.270 & 0.164 & 0.074 & 0.052 & 0.052 & 0.051 & 0.911 & 0.729 & 0.415 \\ 
EL & 0.060 & 0.056 & 0.044 & 0.124 & 0.096 & 0.062 & 0.053 & 0.052 & 0.051 & 0.703 & 0.565 & 0.383 \\ 
\hline 
\multicolumn{3}{l}{\textbf{u-shaped break}} \\ \hline 
$\tau$-SB, $B = T^{0.5}$   & 0.065 & 0.067 & 0.064 & 0.247 & 0.143 & 0.089 & 0.057 & 0.057 & 0.058 & 0.810 & 0.650 & 0.452 \\ 
$\tau$-SB, $B = T^{0.6}$   & 0.066 & 0.070 & 0.070 & 0.271 & 0.135 & 0.072 & 0.059 & 0.060 & 0.062 & 0.918 & 0.740 & 0.464 \\ 
$\tau$-SB, $B = T^{0.7}$ & 0.079 & 0.105 & 0.109 & 0.290 & 0.136 & 0.069 & 0.059 & 0.062 & 0.066 & 0.954 & 0.691 & 0.280 \\ 
$\tau$-SB, $B = T^{0.8}$   & 0.055 & 0.067 & 0.069 & 0.253 & 0.093 & 0.034 & 0.053 & 0.064 & 0.079 & 0.937 & 0.520 & 0.116 \\ 
$\tau$-FB, $B = 0.2T$    & 0.040 & 0.031 & 0.025 & 0.170 & 0.059 & 0.018 & 0.045 & 0.041 & 0.036 & 0.885 & 0.477 & 0.149 \\ 
$\tau$-FB, $B = 0.4T$   & 0.045 & 0.037 & 0.030 & 0.196 & 0.047 & 0.010 & 0.048 & 0.046 & 0.042 & 0.878 & 0.364 & 0.049 \\ 
$\tau$-FB, $B = 0.6T$   & 0.043 & 0.044 & 0.057 & 0.183 & 0.044 & 0.013 & 0.046 & 0.042 & 0.038 & 0.852 & 0.256 & 0.024 \\ 
ADF & 0.046 & 0.027 & 0.011 & 0.181 & 0.030 & 0.002 & 0.048 & 0.041 & 0.030 & 0.915 & 0.329 & 0.018 \\ 
DF-GLS & 0.077 & 0.068 & 0.049 & 0.435 & 0.163 & 0.037 & 0.059 & 0.059 & 0.056 & 0.885 & 0.634 & 0.230 \\ 
DF-GLS-trend & 0.063 & 0.040 & 0.017 & 0.148 & 0.016 & 0.000 & 0.051 & 0.046 & 0.036 & 0.743 & 0.126 & 0.001 \\ 
EL & 0.066 & 0.065 & 0.054 & 0.132 & 0.112 & 0.080 & 0.055 & 0.057 & 0.057 & 0.702 & 0.568 & 0.405 \\  
\hline 
\multicolumn{3}{l}{\textbf{continuous break}} \\  \hline 
$\tau$-SB, $B = T^{0.5}$   & 0.055 & 0.036 & 0.017 & 0.266 & 0.128 & 0.029 & 0.055 & 0.048 & 0.038 & 0.852 & 0.808 & 0.719 \\ 
$\tau$-SB, $B = T^{0.6}$   & 0.055 & 0.035 & 0.016 & 0.300 & 0.123 & 0.019 & 0.056 & 0.048 & 0.038 & 0.950 & 0.911 & 0.789 \\ 
$\tau$-SB, $B = T^{0.7}$ & 0.051 & 0.032 & 0.014 & 0.314 & 0.100 & 0.011 & 0.055 & 0.047 & 0.036 & 0.983 & 0.928 & 0.680 \\ 
$\tau$-SB, $B = T^{0.8}$   & 0.042 & 0.028 & 0.014 & 0.287 & 0.091 & 0.010 & 0.046 & 0.039 & 0.030 & 0.983 & 0.873 & 0.449 \\ 
$\tau$-FB, $B = 0.2T$    & 0.036 & 0.023 & 0.011 & 0.214 & 0.080 & 0.012 & 0.044 & 0.037 & 0.029 & 0.953 & 0.846 & 0.525 \\ 
$\tau$-FB, $B = 0.4T$   & 0.040 & 0.027 & 0.014 & 0.261 & 0.097 & 0.014 & 0.046 & 0.040 & 0.031 & 0.972 & 0.855 & 0.472 \\ 
$\tau$-FB, $B = 0.6T$   & 0.041 & 0.028 & 0.015 & 0.269 & 0.105 & 0.016 & 0.044 & 0.039 & 0.032 & 0.970 & 0.845 & 0.461 \\ 
ADF & 0.045 & 0.027 & 0.010 & 0.151 & 0.011 & 0.000 & 0.048 & 0.040 & 0.029 & 0.895 & 0.235 & 0.003 \\ 
DF-GLS & 0.064 & 0.039 & 0.015 & 0.351 & 0.045 & 0.001 & 0.056 & 0.046 & 0.035 & 0.885 & 0.541 & 0.060 \\ 
DF-GLS-trend & 0.061 & 0.041 & 0.021 & 0.230 & 0.076 & 0.011 & 0.050 & 0.044 & 0.035 & 0.891 & 0.607 & 0.192 \\ 
EL & 0.059 & 0.054 & 0.047 & 0.129 & 0.116 & 0.097 & 0.053 & 0.051 & 0.048 & 0.744 & 0.710 & 0.652 \\  
\hline 
\multicolumn{3}{l}{\textbf{u-shaped break in intercept}} \\  \hline 
$\tau$-SB, $B = T^{0.5}$   & 0.064 & 0.061 & 0.056 & 0.236 & 0.123 & 0.068 & 0.056 & 0.056 & 0.055 & 0.807 & 0.636 & 0.424 \\ 
$\tau$-SB, $B = T^{0.6}$   & 0.065 & 0.064 & 0.058 & 0.254 & 0.109 & 0.049 & 0.059 & 0.059 & 0.058 & 0.915 & 0.718 & 0.414 \\ 
$\tau$-SB, $B = T^{0.7}$ & 0.077 & 0.092 & 0.089 & 0.262 & 0.099 & 0.039 & 0.058 & 0.059 & 0.060 & 0.950 & 0.640 & 0.202 \\ 
$\tau$-SB, $B = T^{0.8}$   & 0.053 & 0.062 & 0.058 & 0.230 & 0.066 & 0.017 & 0.052 & 0.062 & 0.073 & 0.929 & 0.444 & 0.063 \\ 
$\tau$-FB, $B = 0.2T$    & 0.038 & 0.029 & 0.025 & 0.160 & 0.055 & 0.028 & 0.044 & 0.039 & 0.032 & 0.877 & 0.435 & 0.128 \\ 
$\tau$-FB, $B = 0.4T$   & 0.044 & 0.038 & 0.039 & 0.198 & 0.073 & 0.046 & 0.048 & 0.045 & 0.043 & 0.877 & 0.408 & 0.115 \\ 
$\tau$-FB, $B = 0.6T$   & 0.042 & 0.047 & 0.086 & 0.201 & 0.113 & 0.134 & 0.045 & 0.040 & 0.038 & 0.881 & 0.426 & 0.164 \\ 
ADF & 0.043 & 0.022 & 0.007 & 0.112 & 0.004 & 0.000 & 0.047 & 0.037 & 0.025 & 0.784 & 0.051 & 0.000 \\ 
DF-GLS & 0.073 & 0.060 & 0.037 & 0.353 & 0.066 & 0.004 & 0.058 & 0.055 & 0.049 & 0.907 & 0.578 & 0.069 \\ 
DF-GLS-trend & 0.063 & 0.040 & 0.017 & 0.148 & 0.016 & 0.000 & 0.051 & 0.046 & 0.036 & 0.743 & 0.126 & 0.001 \\ 
EL & 0.066 & 0.065 & 0.054 & 0.132 & 0.112 & 0.080 & 0.055 & 0.057 & 0.057 & 0.702 & 0.568 & 0.405 \\   
\hline 
\end{tabular} 
\end{scriptsize}
\hspace*{0.1ex}
\parbox{15.7cm}{ \vspace{0.5ex} 
\tiny
Note: Simulation results are reported for 100,000 replications. 
The errors $u_t$ are simulated independently as standard normal random variables. The series are not pre-whitened ($p=0$).
The rejection frequencies are based on the asymptotic critical values for a significance level of $5\%$.
}
\label{tab_trend1}
\end{table}

\begin{table}[hp]
\caption{Size and power results under different trends and i.i.d.\ errors (2/2)}
\centering
\begin{scriptsize}
\begin{tabular}{l|ccc|ccc|ccc|ccc}
\multicolumn{1}{c|}{sample size} &  \multicolumn{6}{c|}{$T=100$}  &  \multicolumn{6}{c}{$T=300$}  \\ \hline
\multicolumn{1}{c|}{$\rho$} &  \multicolumn{3}{c|}{$\rho = 1$} & \multicolumn{3}{c|}{$\rho = 0.9$} &  \multicolumn{3}{c|}{$\rho = 1$}  &  \multicolumn{3}{c}{$\rho = 0.9$}  \\
\multicolumn{1}{c|}{$\lambda$} & $3$ & $6$ & $9$ & $3$ & $6$ & $9$ & $3$ & $6$ & $9$ & $3$ & $6$ & $9$  \\ 
\hline  
\multicolumn{3}{l}{\textbf{LSTAR break}} \\ \hline
$\tau$-SB, $B = T^{0.5}$   & 0.057 & 0.042 & 0.024 & 0.282 & 0.170 & 0.062 & 0.055 & 0.051 & 0.044 & 0.856 & 0.827 & 0.769 \\ 
$\tau$-SB, $B = T^{0.6}$   & 0.057 & 0.040 & 0.022 & 0.318 & 0.161 & 0.041 & 0.057 & 0.051 & 0.043 & 0.954 & 0.927 & 0.853 \\ 
$\tau$-SB, $B = T^{0.7}$ & 0.054 & 0.037 & 0.019 & 0.327 & 0.118 & 0.017 & 0.056 & 0.049 & 0.040 & 0.985 & 0.945 & 0.771 \\ 
$\tau$-SB, $B = T^{0.8}$   & 0.044 & 0.031 & 0.017 & 0.287 & 0.092 & 0.011 & 0.047 & 0.042 & 0.035 & 0.983 & 0.870 & 0.449 \\ 
$\tau$-FB, $B = 0.2T$    & 0.038 & 0.026 & 0.014 & 0.222 & 0.093 & 0.019 & 0.044 & 0.039 & 0.032 & 0.956 & 0.868 & 0.599 \\ 
$\tau$-FB, $B = 0.4T$   & 0.042 & 0.030 & 0.018 & 0.258 & 0.098 & 0.016 & 0.047 & 0.042 & 0.035 & 0.967 & 0.821 & 0.411 \\ 
$\tau$-FB, $B = 0.6T$   & 0.042 & 0.032 & 0.019 & 0.262 & 0.103 & 0.019 & 0.045 & 0.041 & 0.033 & 0.964 & 0.799 & 0.377 \\ 
ADF & 0.049 & 0.034 & 0.019 & 0.189 & 0.028 & 0.001 & 0.049 & 0.044 & 0.036 & 0.932 & 0.402 & 0.019 \\ 
DF-GLS & 0.070 & 0.051 & 0.029 & 0.415 & 0.101 & 0.006 & 0.056 & 0.051 & 0.043 & 0.899 & 0.671 & 0.197 \\ 
DF-GLS-trend & 0.063 & 0.050 & 0.033 & 0.265 & 0.142 & 0.048 & 0.051 & 0.046 & 0.041 & 0.916 & 0.758 & 0.449 \\ 
EL & 0.059 & 0.053 & 0.046 & 0.129 & 0.115 & 0.094 & 0.053 & 0.051 & 0.048 & 0.741 & 0.704 & 0.644 \\ 
\hline 
\multicolumn{3}{l}{\textbf{offsetting LSTAR break}} \\ \hline
$\tau$-SB, $B = T^{0.5}$   & 0.056 & 0.038 & 0.019 & 0.276 & 0.152 & 0.048 & 0.056 & 0.049 & 0.041 & 0.854 & 0.819 & 0.746 \\ 
$\tau$-SB, $B = T^{0.6}$   & 0.055 & 0.036 & 0.017 & 0.307 & 0.142 & 0.032 & 0.057 & 0.049 & 0.039 & 0.952 & 0.916 & 0.813 \\ 
$\tau$-SB, $B = T^{0.7}$ & 0.052 & 0.033 & 0.015 & 0.320 & 0.115 & 0.016 & 0.056 & 0.048 & 0.037 & 0.983 & 0.925 & 0.671 \\ 
$\tau$-SB, $B = T^{0.8}$   & 0.042 & 0.027 & 0.013 & 0.281 & 0.088 & 0.011 & 0.047 & 0.040 & 0.031 & 0.978 & 0.809 & 0.326 \\ 
$\tau$-FB, $B = 0.2T$    & 0.036 & 0.023 & 0.011 & 0.212 & 0.081 & 0.014 & 0.043 & 0.038 & 0.029 & 0.950 & 0.823 & 0.471 \\ 
$\tau$-FB, $B = 0.4T$   & 0.040 & 0.026 & 0.012 & 0.240 & 0.077 & 0.010 & 0.046 & 0.039 & 0.031 & 0.949 & 0.691 & 0.225 \\ 
$\tau$-FB, $B = 0.6T$   & 0.039 & 0.025 & 0.012 & 0.229 & 0.062 & 0.006 & 0.045 & 0.039 & 0.030 & 0.930 & 0.573 & 0.115 \\ 
ADF & 0.052 & 0.048 & 0.048 & 0.269 & 0.135 & 0.059 & 0.050 & 0.047 & 0.045 & 0.981 & 0.837 & 0.452 \\ 
DF-GLS & 0.069 & 0.045 & 0.023 & 0.435 & 0.136 & 0.015 & 0.055 & 0.048 & 0.039 & 0.845 & 0.511 & 0.142 \\ 
DF-GLS-trend & 0.060 & 0.038 & 0.018 & 0.211 & 0.054 & 0.005 & 0.049 & 0.042 & 0.033 & 0.854 & 0.458 & 0.074 \\ 
EL & 0.060 & 0.055 & 0.049 & 0.131 & 0.121 & 0.106 & 0.053 & 0.052 & 0.049 & 0.747 & 0.723 & 0.684 \\ 
\hline 
\multicolumn{3}{l}{\textbf{triangular break}} \\ \hline
$\tau$-SB, $B = T^{0.5}$   & 0.055 & 0.040 & 0.023 & 0.282 & 0.168 & 0.060 & 0.055 & 0.050 & 0.042 & 0.855 & 0.824 & 0.761 \\ 
$\tau$-SB, $B = T^{0.6}$   & 0.056 & 0.039 & 0.021 & 0.318 & 0.164 & 0.045 & 0.057 & 0.050 & 0.041 & 0.954 & 0.924 & 0.847 \\ 
$\tau$-SB, $B = T^{0.7}$ & 0.054 & 0.036 & 0.019 & 0.335 & 0.142 & 0.029 & 0.056 & 0.050 & 0.040 & 0.985 & 0.947 & 0.769 \\ 
$\tau$-SB, $B = T^{0.8}$   & 0.042 & 0.028 & 0.015 & 0.290 & 0.105 & 0.017 & 0.046 & 0.041 & 0.034 & 0.977 & 0.826 & 0.388 \\ 
$\tau$-FB, $B = 0.2T$    & 0.037 & 0.026 & 0.014 & 0.224 & 0.100 & 0.024 & 0.044 & 0.039 & 0.031 & 0.955 & 0.864 & 0.579 \\ 
$\tau$-FB, $B = 0.4T$   & 0.041 & 0.028 & 0.014 & 0.258 & 0.098 & 0.018 & 0.047 & 0.041 & 0.032 & 0.949 & 0.715 & 0.273 \\ 
$\tau$-FB, $B = 0.6T$   & 0.041 & 0.028 & 0.016 & 0.262 & 0.105 & 0.021 & 0.045 & 0.040 & 0.032 & 0.957 & 0.758 & 0.333 \\ 
ADF & 0.052 & 0.047 & 0.042 & 0.256 & 0.105 & 0.027 & 0.051 & 0.048 & 0.045 & 0.975 & 0.782 & 0.331 \\ 
DF-GLS & 0.067 & 0.045 & 0.023 & 0.459 & 0.175 & 0.027 & 0.056 & 0.049 & 0.039 & 0.891 & 0.682 & 0.314 \\ 
DF-GLS-trend & 0.059 & 0.038 & 0.018 & 0.202 & 0.048 & 0.004 & 0.050 & 0.042 & 0.033 & 0.841 & 0.409 & 0.052 \\ 
EL & 0.060 & 0.058 & 0.054 & 0.133 & 0.127 & 0.118 & 0.053 & 0.052 & 0.051 & 0.752 & 0.742 & 0.726 \\ 
\hline 
\multicolumn{3}{l}{\textbf{Fourier break}} \\ \hline
$\tau$-SB, $B = T^{0.5}$   & 0.054 & 0.034 & 0.015 & 0.261 & 0.119 & 0.025 & 0.055 & 0.048 & 0.038 & 0.852 & 0.809 & 0.718 \\ 
$\tau$-SB, $B = T^{0.6}$   & 0.054 & 0.033 & 0.014 & 0.287 & 0.103 & 0.013 & 0.056 & 0.048 & 0.037 & 0.951 & 0.905 & 0.762 \\ 
$\tau$-SB, $B = T^{0.7}$ & 0.050 & 0.028 & 0.011 & 0.289 & 0.074 & 0.006 & 0.055 & 0.045 & 0.034 & 0.981 & 0.893 & 0.496 \\ 
$\tau$-SB, $B = T^{0.8}$   & 0.040 & 0.022 & 0.009 & 0.247 & 0.051 & 0.003 & 0.046 & 0.038 & 0.028 & 0.963 & 0.644 & 0.113 \\ 
$\tau$-FB, $B = 0.2T$    & 0.035 & 0.020 & 0.008 & 0.195 & 0.056 & 0.006 & 0.043 & 0.036 & 0.027 & 0.944 & 0.756 & 0.292 \\ 
$\tau$-FB, $B = 0.4T$   & 0.038 & 0.021 & 0.009 & 0.218 & 0.053 & 0.004 & 0.045 & 0.037 & 0.027 & 0.923 & 0.526 & 0.079 \\ 
$\tau$-FB, $B = 0.6T$   & 0.038 & 0.022 & 0.009 & 0.223 & 0.055 & 0.004 & 0.044 & 0.037 & 0.027 & 0.930 & 0.557 & 0.095 \\ 
ADF & 0.048 & 0.037 & 0.026 & 0.217 & 0.054 & 0.007 & 0.049 & 0.044 & 0.036 & 0.959 & 0.594 & 0.102 \\ 
DF-GLS & 0.066 & 0.037 & 0.015 & 0.427 & 0.115 & 0.009 & 0.055 & 0.046 & 0.035 & 0.885 & 0.633 & 0.205 \\ 
DF-GLS-trend & 0.057 & 0.031 & 0.011 & 0.172 & 0.023 & 0.001 & 0.048 & 0.039 & 0.028 & 0.808 & 0.267 & 0.010 \\ 
EL & 0.061 & 0.061 & 0.061 & 0.134 & 0.134 & 0.134 & 0.053 & 0.053 & 0.053 & 0.755 & 0.755 & 0.755 \\ 
\hline 
\end{tabular} 
\end{scriptsize}
\hspace*{0.1ex}
\parbox{15.7cm}{ \vspace{0.5ex} 
\tiny
Note: Simulation results are reported for 100,000 replications. 
The errors $u_t$ are simulated independently as standard normal random variables. The series are not pre-whitened ($p=0$).
The rejection frequencies are based on the asymptotic critical values for a significance level of $5\%$.
}
\label{tab_trend2}
\end{table}

\begin{table}[hp]
\caption{Size and power results under different trends and AR(1) errors}
\centering
\begin{scriptsize}
\begin{tabular}{l|ccc|ccc|ccc|ccc}
\multicolumn{1}{c|}{sample size} &  \multicolumn{6}{c|}{$T=100$}  &  \multicolumn{6}{c}{$T=300$}  \\ \hline
\multicolumn{1}{c|}{$\rho$} &  \multicolumn{3}{c|}{$\rho = 1$} & \multicolumn{3}{c|}{$\rho = 0.9$} &  \multicolumn{3}{c|}{$\rho = 1$}  &  \multicolumn{3}{c}{$\rho = 0.9$}  \\
\multicolumn{1}{c|}{$\lambda$} & $3$ & $6$ & $9$ &  $3$ & $6$ & $9$ &  $3$ & $6$ & $9$ &  $3$ & $6$ & $9$  \\ 
\hline 
\multicolumn{3}{l}{\textbf{sharp break}} \\ \hline
$\tau\text{-SB}_p$, $B = T^{0.5}$   & 0.006 & 0.007 & 0.009 & 0.069 & 0.042 & 0.032 & 0.015 & 0.014 & 0.012 & 0.607 & 0.477 & 0.362 \\ 
$\tau\text{-SB}_p$, $B = T^{0.6}$   & 0.019 & 0.022 & 0.028 & 0.160 & 0.111 & 0.087 & 0.033 & 0.032 & 0.030 & 0.837 & 0.754 & 0.650 \\ 
$\tau\text{-SB}_p$, $B = T^{0.7}$    & 0.034 & 0.043 & 0.060 & 0.250 & 0.187 & 0.147 & 0.044 & 0.044 & 0.046 & 0.934 & 0.877 & 0.781 \\ 
$\tau\text{-SB}_p$, $B = T^{0.8}$   & 0.051 & 0.108 & 0.172 & 0.307 & 0.325 & 0.309 & 0.043 & 0.047 & 0.057 & 0.946 & 0.860 & 0.712 \\ 
$\tau\text{-FB}_p$, $B = 0.2T$    & 0.023 & 0.045 & 0.069 & 0.149 & 0.167 & 0.156 & 0.038 & 0.043 & 0.058 & 0.882 & 0.790 & 0.692 \\ 
$\tau\text{-FB}_p$, $B = 0.4T$   & 0.029 & 0.024 & 0.019 & 0.187 & 0.092 & 0.039 & 0.042 & 0.037 & 0.033 & 0.920 & 0.779 & 0.555 \\ 
$\tau\text{-FB}_p$, $B = 0.6T$   & 0.029 & 0.016 & 0.008 & 0.192 & 0.072 & 0.016 & 0.041 & 0.033 & 0.024 & 0.918 & 0.767 & 0.501 \\  
\hline 
\multicolumn{3}{l}{\textbf{u-shaped break}} \\ \hline
$\tau\text{-SB}_p$, $B = T^{0.5}$   & 0.007 & 0.011 & 0.019 & 0.079 & 0.066 & 0.060 & 0.015 & 0.014 & 0.013 & 0.610 & 0.471 & 0.346 \\ 
$\tau\text{-SB}_p$, $B = T^{0.6}$   & 0.020 & 0.032 & 0.047 & 0.175 & 0.148 & 0.129 & 0.032 & 0.032 & 0.035 & 0.833 & 0.726 & 0.592 \\ 
$\tau\text{-SB}_p$, $B = T^{0.7}$    & 0.051 & 0.097 & 0.140 & 0.305 & 0.293 & 0.253 & 0.045 & 0.049 & 0.058 & 0.926 & 0.831 & 0.680 \\ 
$\tau\text{-SB}_p$, $B = T^{0.8}$   & 0.034 & 0.053 & 0.080 & 0.254 & 0.213 & 0.174 & 0.044 & 0.055 & 0.072 & 0.935 & 0.816 & 0.623 \\ 
$\tau\text{-FB}_p$, $B = 0.2T$    & 0.024 & 0.044 & 0.056 & 0.149 & 0.145 & 0.117 & 0.037 & 0.042 & 0.054 & 0.862 & 0.694 & 0.515 \\ 
$\tau\text{-FB}_p$, $B = 0.4T$   & 0.034 & 0.048 & 0.061 & 0.211 & 0.150 & 0.103 & 0.043 & 0.045 & 0.053 & 0.902 & 0.701 & 0.435 \\ 
$\tau\text{-FB}_p$, $B = 0.6T$   & 0.037 & 0.060 & 0.089 & 0.218 & 0.164 & 0.127 & 0.043 & 0.048 & 0.060 & 0.896 & 0.662 & 0.380 \\
\hline 
\multicolumn{3}{l}{\textbf{continuous break}} \\  \hline
$\tau\text{-SB}_p$, $B = T^{0.5}$   & 0.006 & 0.005 & 0.004 & 0.079 & 0.049 & 0.022 & 0.016 & 0.015 & 0.015 & 0.654 & 0.594 & 0.511 \\ 
$\tau\text{-SB}_p$, $B = T^{0.6}$   & 0.018 & 0.016 & 0.013 & 0.173 & 0.115 & 0.060 & 0.033 & 0.032 & 0.030 & 0.859 & 0.818 & 0.746 \\ 
$\tau\text{-SB}_p$, $B = T^{0.7}$    & 0.030 & 0.027 & 0.023 & 0.256 & 0.173 & 0.087 & 0.044 & 0.043 & 0.040 & 0.943 & 0.906 & 0.826 \\ 
$\tau\text{-SB}_p$, $B = T^{0.8}$   & 0.029 & 0.027 & 0.023 & 0.250 & 0.172 & 0.092 & 0.041 & 0.039 & 0.037 & 0.956 & 0.908 & 0.792 \\ 
$\tau\text{-FB}_p$, $B = 0.2T$    & 0.019 & 0.017 & 0.015 & 0.151 & 0.107 & 0.060 & 0.037 & 0.036 & 0.034 & 0.902 & 0.851 & 0.751 \\ 
$\tau\text{-FB}_p$, $B = 0.4T$   & 0.032 & 0.029 & 0.025 & 0.230 & 0.168 & 0.100 & 0.043 & 0.041 & 0.040 & 0.943 & 0.897 & 0.796 \\ 
$\tau\text{-FB}_p$, $B = 0.6T$   & 0.033 & 0.031 & 0.027 & 0.237 & 0.177 & 0.108 & 0.043 & 0.041 & 0.039 & 0.943 & 0.898 & 0.792 \\   
\hline 
\multicolumn{3}{l}{\textbf{LSTAR break}} \\ \hline
$\tau\text{-SB}_p$, $B = T^{0.5}$   & 0.005 & 0.004 & 0.003 & 0.063 & 0.025 & 0.010 & 0.016 & 0.015 & 0.014 & 0.613 & 0.505 & 0.424 \\ 
$\tau\text{-SB}_p$, $B = T^{0.6}$   & 0.017 & 0.014 & 0.010 & 0.149 & 0.078 & 0.035 & 0.033 & 0.033 & 0.030 & 0.842 & 0.779 & 0.706 \\ 
$\tau\text{-SB}_p$, $B = T^{0.7}$    & 0.030 & 0.025 & 0.019 & 0.231 & 0.126 & 0.055 & 0.043 & 0.042 & 0.039 & 0.937 & 0.890 & 0.809 \\ 
$\tau\text{-SB}_p$, $B = T^{0.8}$   & 0.028 & 0.024 & 0.018 & 0.222 & 0.120 & 0.053 & 0.041 & 0.039 & 0.037 & 0.951 & 0.887 & 0.746 \\ 
$\tau\text{-FB}_p$, $B = 0.2T$    & 0.018 & 0.015 & 0.011 & 0.133 & 0.074 & 0.037 & 0.037 & 0.036 & 0.034 & 0.894 & 0.829 & 0.723 \\ 
$\tau\text{-FB}_p$, $B = 0.4T$   & 0.031 & 0.027 & 0.021 & 0.204 & 0.123 & 0.063 & 0.044 & 0.042 & 0.040 & 0.936 & 0.866 & 0.729 \\ 
$\tau\text{-FB}_p$, $B = 0.6T$   & 0.032 & 0.028 & 0.023 & 0.215 & 0.134 & 0.071 & 0.042 & 0.041 & 0.039 & 0.934 & 0.858 & 0.717 \\  
\hline 
\multicolumn{3}{l}{\textbf{Fourier break}} \\ \hline
$\tau\text{-SB}_p$, $B = T^{0.5}$   & 0.005 & 0.005 & 0.003 & 0.081 & 0.055 & 0.029 & 0.016 & 0.015 & 0.015 & 0.658 & 0.608 & 0.533 \\ 
$\tau\text{-SB}_p$, $B = T^{0.6}$   & 0.017 & 0.014 & 0.011 & 0.176 & 0.122 & 0.067 & 0.033 & 0.032 & 0.030 & 0.860 & 0.821 & 0.750 \\ 
$\tau\text{-SB}_p$, $B = T^{0.7}$    & 0.030 & 0.025 & 0.019 & 0.257 & 0.175 & 0.090 & 0.044 & 0.042 & 0.039 & 0.941 & 0.897 & 0.796 \\ 
$\tau\text{-SB}_p$, $B = T^{0.8}$   & 0.029 & 0.024 & 0.018 & 0.247 & 0.162 & 0.079 & 0.041 & 0.039 & 0.036 & 0.946 & 0.853 & 0.648 \\ 
$\tau\text{-FB}_p$, $B = 0.2T$    & 0.019 & 0.016 & 0.012 & 0.150 & 0.105 & 0.058 & 0.037 & 0.036 & 0.033 & 0.898 & 0.832 & 0.693 \\ 
$\tau\text{-FB}_p$, $B = 0.4T$   & 0.031 & 0.026 & 0.021 & 0.224 & 0.153 & 0.082 & 0.043 & 0.041 & 0.038 & 0.925 & 0.811 & 0.593 \\ 
$\tau\text{-FB}_p$, $B = 0.6T$   & 0.032 & 0.028 & 0.021 & 0.231 & 0.158 & 0.086 & 0.043 & 0.041 & 0.038 & 0.930 & 0.828 & 0.621 \\    
\hline
\end{tabular} 
\end{scriptsize}
\parbox{15.9cm}{ \vspace{0.5ex} 
\tiny
Note: Simulation results are reported for 100,000 replications. 
	The errors $u_t$ are simulated from $u_t = 0.5 u_{t-1} + \epsilon_t$ with independent standard normal innovations, and the series are pre-whitened with a lag order $p$ that is determined from the BIC.
	The rejection frequencies are based on the asymptotic critical values for a significance level of $5\%$.
}
\label{tab_trends_ar}
\end{table}

Tables \ref{tab_notrend}--\ref{tab_HCtrends} present size and actual power results under different model specifications.
For smaller sample sizes, the pooled tests have small size distortions, which become larger as the break gets larger.
However, for larger sample sizes, the size distortions decline. 
Overall, the size levels are similar to those obtained from using the conventional unit root tests.

The power of the pooled tests depends on the blocklength.
In case of no break, a larger blocklength implies higher power results, which is in line with the theoretical findings that those tests have power in a $1/\sqrt{BT}$ neighborhood of the unit root hypothesis.
For blocklengths of $B=T^{0.8}$ in the small-$b$ case and $B = 0.6 T$ in the fixed-$b$ case, the power results are similar to those from the ADF test and the Dickey-Fuller GLS test, where the ordering depends on the initial condition (cf.\ Figure \ref{fig_initial}).
Hence, none of the tests dominates the pooled tests uniformly across these small-sample specifications (although, asymptotically, those tests have power in a $1/T$ neighborhood of the unit root hypothesis).
Furthermore, smaller blocklengths, such as $T^{0.6}$ in the small-$b$ context and $0.2 T$ in the fixed-$b$ context, still yield reasonably high power.
In particular, the EL test performs much worse in all cases.
The size and power results obtained under the AR(1) error specification with both fixed and flexible lag augmentation for the pre-whitening scheme are similar to those produced by i.i.d.\ errors.

\begin{table}[ht]
\caption{Size and power results of robust tests under breaks in trend and variance
}
\centering
\begin{scriptsize}
\begin{tabular}{l|ccc|ccc|ccc|ccc}
\multicolumn{1}{c|}{sample size} & \multicolumn{6}{c|}{$T=100$} & \multicolumn{6}{c}{$T=300$} \\ \hline
\multicolumn{1}{c|}{$\rho$} & \multicolumn{3}{c|}{$\rho = 1$} & \multicolumn{3}{c|}{$\rho = 0.9$} & \multicolumn{3}{c|}{$\rho = 1$} & \multicolumn{3}{c}{$\rho = 0.9$} \\
\multicolumn{1}{c|}{$\lambda$} & $2$ & $3$ & $4$ & $2$ & $3$ & $4$ & $2$ & $3$ & $4$ & $2$ & $3$ & $4$ \\
\hline
\multicolumn{5}{l}{\textbf{sharp break in variance}} \\ \hline
$\tau$-SB, $B = T^{0.5}$   & 0.067 & 0.069 & 0.069 & 0.344 & 0.337 & 0.329 & 0.057 & 0.056 & 0.056 & 0.847 & 0.806 & 0.767 \\ 
$\tau$-SB, $B = T^{0.6}$   & 0.071 & 0.075 & 0.077 & 0.420 & 0.421 & 0.416 & 0.062 & 0.061 & 0.061 & 0.954 & 0.933 & 0.909 \\ 
$\tau$-SB, $B = T^{0.7}$ & 0.081 & 0.095 & 0.107 & 0.526 & 0.565 & 0.585 & 0.068 & 0.072 & 0.074 & 0.992 & 0.987 & 0.981 \\ 
$\tau$-SB, $B = T^{0.8}$   & 0.085 & 0.124 & 0.162 & 0.569 & 0.683 & 0.756 & 0.082 & 0.116 & 0.147 & 0.999 & 1.000 & 1.000 \\ 
$\tau$-FB, $B = 0.2T$    & 0.040 & 0.039 & 0.040 & 0.283 & 0.261 & 0.238 & 0.045 & 0.044 & 0.042 & 0.947 & 0.882 & 0.812 \\ 
$\tau$-FB, $B = 0.4T$   & 0.043 & 0.042 & 0.041 & 0.346 & 0.308 & 0.276 & 0.047 & 0.046 & 0.045 & 0.982 & 0.935 & 0.876 \\ 
$\tau$-FB, $B = 0.6T$   & 0.042 & 0.040 & 0.042 & 0.349 & 0.327 & 0.307 & 0.045 & 0.045 & 0.045 & 0.989 & 0.974 & 0.947 \\
\hline
\multicolumn{5}{l}{\textbf{sharp break in trend and variance}} \\ \hline
$\tau$-SB, $B = T^{0.5}$   & 0.066 & 0.068 & 0.070 & 0.324 & 0.305 & 0.283 & 0.057 & 0.056 & 0.056 & 0.836 & 0.786 & 0.737 \\ 
$\tau$-SB, $B = T^{0.6}$   & 0.071 & 0.074 & 0.076 & 0.391 & 0.370 & 0.344 & 0.062 & 0.061 & 0.061 & 0.946 & 0.916 & 0.881 \\ 
$\tau$-SB, $B = T^{0.7}$ & 0.080 & 0.091 & 0.099 & 0.474 & 0.470 & 0.444 & 0.067 & 0.071 & 0.073 & 0.988 & 0.976 & 0.959 \\ 
$\tau$-SB, $B = T^{0.8}$   & 0.095 & 0.145 & 0.194 & 0.526 & 0.595 & 0.627 & 0.081 & 0.111 & 0.136 & 0.997 & 0.996 & 0.993 \\ 
$\tau$-FB, $B = 0.2T$    & 0.040 & 0.039 & 0.038 & 0.260 & 0.228 & 0.197 & 0.046 & 0.044 & 0.043 & 0.935 & 0.854 & 0.765 \\ 
$\tau$-FB, $B = 0.4T$   & 0.044 & 0.042 & 0.043 & 0.295 & 0.240 & 0.200 & 0.047 & 0.046 & 0.046 & 0.960 & 0.872 & 0.772 \\ 
$\tau$-FB, $B = 0.6T$   & 0.042 & 0.043 & 0.047 & 0.292 & 0.240 & 0.205 & 0.046 & 0.045 & 0.046 & 0.954 & 0.866 & 0.766 \\  
\hline
\end{tabular} 
\end{scriptsize}
\parbox{15.7cm}{ \vspace{0.5ex} 
\tiny
Note: Simulation results are reported for 100,000 replications. 
	The errors $u_t$ are simulated independently as standard normal random variables, and the series are not pre-whitened ($p=0$).
	The sharp break specification is defined by a break in the variance at $2/3$ of the sample.
	The rejection frequencies are based on the asymptotic critical values for a significance level of $5\%$.
}
\label{tab_HCtrends}
\end{table}

As the tests are designed to yield higher power in the presence of slowly varying trends and breaks, we compare the size-adjusted powers of the tests under the trend specifications presented in Table \ref{tab_trends} and Figure \ref{fig_trends}.
For large break sizes $\lambda$, it is shown that the smaller the blocklength, the greater the power results.
In most cases, the pooled tests have greater power than the ADF, the DF-GLS, the DF-GLS-trend, and the EL test.
Furthermore, the power results of the pooled tests are quite uniform across different trend specifications when compared to those of the conventional tests.

Table \ref{tab_trends_ar} shows that the pooled tests have reasonable size and power properties under the presence of AR(1) errors and different trend specifications.
Furthermore, from Table \ref{tab_HCtrends}, we can conclude that the tests are sized correctly and have good power properties in the presence of a break in the variance and in the trend function.

The blocklength $B$ is a tuning parameter that needs to be chosen carefully, and any optimality result would depend on the actual trend model.
In practice, however, the trend model is unknown, which makes it hard to derive an optimal blocklength.
Although theoretical recommendations cannot be formulated based on the current analysis, the small-$b$ tests with $B = T^{0.7}$ and the fixed-$b$ tests with $T = 0.2 B$ yield very promising results for all trend functions studied in this paper and are therefore recommended as the default settings.

\section{Conclusion}	\label{sec_conclusion1}

We have presented two variants of a unit root test under an unknown trend specification that are robust under both heteroskedasticity and autocorrelation.
When applied to finite samples, the tests show good size properties.
The fixed-$b$ pooled test statistic converges to a functional of a Brownian motion under the unit root hypothesis, while the small-$b$ variant shows a standard normal distribution in the limit.
Autocorrelation-robust versions of the tests were introduced using a pre-whitening scheme.
Monte Carlo simulations indicate that, while under the zero-trend specification, the fixed-$b$ and small-$b$ tests perform similar to the conventional tests in terms of size and power, under sharp breaks as well as smooth changes in the trend, their power is much higher.
Furthermore, the powers of the tests are less sensitive to the initial value when compared to the augmented Dickey-Fuller test and the Dickey-Fuller GLS test.

\section*{Acknowledgements}

I would like to thank Jörg Breitung and Matei Demetrescu for their extensive advice and support. My thanks also go to Hans Manner, Markus Kösler, Robinson Kruse-Becher, Dominik Wied, Nazarii Salish, Uwe Hassler, Martin Wagner, the Co-Editor, and two anonymous referees for their helpful comments. The suggestions made by participants attending the 2015 RMSE meeting in Cologne, the SMYE conference 2017 in Halle (Saale), the SNDE conference 2017 in Paris, and the IAAE conference 2017 in Sapporo are also highly appreciated.
Furthermore, the usage of the CHEOPS HPC cluster for parallel computing and a conference grant of the International Association for Applied Econometrics are greatfully acknowledged.

\section*{Supporting Information}
An accompanying \texttt{R}-package for the application of the tests proposed in this article is available online at \texttt{https://github.com/ottosven/urtrend}.

\newpage
\addcontentsline{toc}{section}{References}
\bibliography{references}
\newpage

\appendix
\addcontentsline{toc}{section}{Appendix}
\section*{Appendix: Proofs}
\renewcommand{\thesection}{A}
\setcounter{lemma}{0} \renewcommand{\thelemma}{A.\arabic{lemma}}

\setcounter{equation}{0} \renewcommand{\theequation}{A.\arabic{equation}}

\subsection{Auxiliary results}

\begin{lemma}	\label{lem_blockfilter_aux}
Let $\rho = 1 - c/\sqrt{BT}$ with $c \geq 0$, let $d_t$ satisfy Assumption \ref{ass:trend}, and let $u_t$ satisfy Assumption \ref{ass:errors}. Furthermore, let $1 \leq s \leq B$. Then, 
\begin{itemize}
	\item[(a)] $\sum_{r=1}^B \big| \sum_{j=1}^{T-B} \Delta d_{r+j} \Delta d_{s+j} \big| = O(1)$
	\item[(b)] $\sum_{r=1}^B \big| \sum_{j=1}^{T-B} \Delta d_{r+j} \Delta x_{s+j} \big| = O_P(T^{1/2})$
\end{itemize}
\end{lemma}

\if1\supplement
{
  \begin{proof}
	Since $d(r)$ is piecewise Lipschitz continuous on the unit interval, there are a finite number of points where $d(r)$ is not continuous. 
Let those points be given by $\{\pi_1, \ldots, \pi_L\}$, where $L < \infty$ and $0 < \pi_1 < \ldots < \pi_L < 1$.
We can represent $d(r)$ by some function $\delta(r)$ that is Lipschitz continuous on the entire domain.
Then, for any $r \in [0,1]$, we obtain $d(r) = \delta(r) + \sum_{l=1}^L \lambda_l 1_{\{ r \geq \pi_l \}}$, with $\sum_{l=1}^L |\lambda_l| < \infty$.
Let $p_l = \lfloor \pi_l T \rfloor$ for $l = 1, \ldots, L$, and let $\delta_t = \delta(t/T)$ for $t = 1, \ldots, T$.
Then, $d_t = \delta_t + \sum_{l=1}^L \lambda_l 1_{\{ t \leq p_l \}}$, and consequently, $\Delta d_{t} = \Delta \delta_{t} + \sum_{l=1}^L \lambda_l 1_{\{ t = p_l\}}$.
Due to the Lipschitz continuity of $\delta(r)$, there exists a constant $C_1 < \infty$, such that
\begin{align}
	\big|\Delta d_{t}\big| \leq C_1 T^{-1} + \sum_{l=1}^L |\lambda_l| 1_{\{ t = p_l\}}, \label{eq:lipschitz-inequality}
\end{align}
for all indices $t=2, \ldots, T$. 
Furthermore, $\sum_{l=1}^L |\lambda_l| = C_2$ and $\sigma(r) < C_3$ for some constants $C_2,C_3 < \infty$, and $C = \max\{ C_1, C_2, C_3 , E[|x_0|], c, 1 \} < \infty$.
For (a), we have
\begin{align*}
	\sum_{r=1}^B \Big| \sum_{j=1}^{T-B} \Delta d_{r+j} \Delta d_{s+j} \Big| \leq 3 C^2 B T^{-1} + \sum_{r=1}^B \sum_{j=1}^{T-B} \sum_{l_1, l_2 = 1}^L \big| \lambda_{l_1} \lambda_{l_2} \big| 1_{\{ r+j=p_{l_1} \}} 1_{\{ s+j=p_{l_2} \}} = O(1).
\end{align*}
To show (b), note that $\Delta x_t = (\rho - 1) x_{t-1} + u_t = (\rho - 1) (x_{t-1} - x_0) + (\rho-1) x_0 + u_t$.
We decompose $\sum_{r=1}^B \Big| \sum_{j=1}^{T-B} \Delta d_{r+j} \Delta x_{s+j} \Big| \leq A_1 + A_2 + A_3$, where
\begin{align*}
	A_1 &= \frac{c \sum_{r=1}^B | \sum_{j=1}^{T-B} \Delta d_{r+j} (x_{s+j-1} - x_0) |}{B^{1/2} T^{1/2}}, \quad
	A_2 = \sum_{r=1}^B \Big| \sum_{j=1}^{T-B} \Delta d_{r+j} u_{s+j} \Big|, \\
	A_3 &= \frac{c \sum_{r=1}^B \sum_{j=1}^{T-B} | \Delta d_{r+j} x_0|}{B^{1/2} T^{1/2}}.
\end{align*}
From the MA-representation $x_t-x_0 = \sum_{m=0}^{t-1} \rho^m u_{t-m}$, inequality \eqref{eq:lipschitz-inequality}, and Jensen's inequality, it follows that
\begin{align*}
&E[|A_1|] 
	\leq (1-\rho) \sum_{r=1}^B \sqrt{ E \bigg[ \Big( \sum_{j=1}^{T-B} \sum_{m=0}^{s+j-2} \rho^m \Delta d_{r+j} u_{s+j-m-1} \Big)^2 \bigg] } \\
	&\leq (1-\rho) \sum_{r=1}^B \sqrt{ \sum_{j=1}^{T-B} \sum_{m_1,m_2=0}^\infty C^2 \rho^{m_1+m_2} |\Delta d_{r+j} \Delta d_{r+j-m_1+m_2}| } \\
	&\leq 2 C^2 (1-\rho) \sqrt{\sum_{j=1}^{T-B} \sum_{m_1,m_2=0}^\infty \rho^{m_1+m_2} } = O(T^{1/2}),
\end{align*}
\begin{align*}
	E[|A_2|]	
	&\leq \sum_{r=1}^B \sqrt{E \Big[ \Big( \sum_{j=1}^{T-B} \Delta d_{r+j} u_{s+j} \Big)^2 \Big]}
	\leq \sum_{r=1}^B \sqrt{ C^2 \sum_{j=1}^{T-B} |\Delta d_{r+j}|^2  } = \sqrt{4C^4 T} = O(T^{1/2}),
\end{align*}
and
\begin{align*}
	E[|A_3|] 
	&= E \bigg[ (1-\rho) \sum_{r=1}^B \Big| \sum_{j=1}^{T-B} \Delta d_{r+j} x_0 \Big| \bigg]
	\leq \frac{c E[|x_0|]}{ B^{1/2} T^{1/2} } \sum_{r=1}^B  \sum_{j=1}^{T-B} | \Delta d_{r+j} |
	\leq \frac{2 C^3 B^{1/2}}{T^{1/2}} = O(1).
\end{align*}
The assertion follows by Markov's inequality and the triangle inequality.
\end{proof}
} \fi

\if0\supplement
{
 	\noindent
  	The proof is available in the supplementary material.
  
} \fi

\subsection{Proof of Lemma \ref{lem_blockfilter1}}

First, we reformulate the numerator and denominator statistics.
Note that
\begin{align*}
	&\Delta y_{t+j} (y_{t+j-1} - y_j) - \Delta x_{t+j} (x_{t+j-1} - x_j) \\
	&= \Delta d_{t+j} (d_{t+j-1} - d_j) + \Delta d_{t+j} (x_{t+j-1} - x_j) + \Delta x_{t+j} (d_{t+j-1} - d_j),
\end{align*}
and
\begin{align*}
	(y_{t+j-1} - y_j)^2 - (x_{t+j-1} - x_j)^2 = (d_{t+j-1} - d_j)^2 + 2(x_{t+j-1} - x_j)(d_{t+j-1} - d_j).
\end{align*}
We decompose $\mathcal Y_{1,T} - \mathcal X_{1,T} = S_1 + S_2 + S_3$ and $\mathcal Y_{2,T} - \mathcal X_{2,T} = S_4 + S_5$, where
\begin{align*}
	S_1 &= \frac{\sum_{j=1}^{T-B} \sum_{t=2}^B \Delta d_{t+j} (d_{t+j-1} - d_j)}{B^{3/2} T^{1/2}}, \quad S_2 = \frac{\sum_{j=1}^{T-B} \sum_{t=2}^B \Delta d_{t+j} (x_{t+j-1} - x_j)}{B^{3/2} T^{1/2}}, \\
	S_3 &= \frac{\sum_{j=1}^{T-B} \sum_{t=2}^B \Delta x_{t+j} (d_{t+j-1} - d_j)}{B^{3/2} T^{1/2}}, \quad S_4 = \frac{\sum_{j=1}^{T-B} \sum_{t=2}^B (d_{t+j-1} - d_j)^2}{B^2 T}, \\
	S_5 &= \frac{\sum_{j=1}^{T-B} \sum_{t=2}^B 2(x_{t+j-1} - x_j)(d_{t+j-1} - d_j)}{B^2 T}.
\end{align*}
Lemma \ref{lem_blockfilter_aux} yields
$S_1 + S_2 + S_3 = O_P(B^{-1/2})$, and $S_4 + S_5 = O_P(T^{-1/2})$,
and the assertion follows by Slutsky's theorem.

\subsection{Proof of Lemma \ref{lem_NumDen} }

\if1\supplement
{
  \textit{(a):}
From the representation $\Delta x_{t+j} = u_{t+j} + \phi x_{t+j-1}$ with $\phi = -c/\sqrt{BT}$, we decompose the numerator statistic into $\mathcal X_{1,T} = S_1 + S_2 + S_3 + S_4$, where
\begin{align*}
	S_1 &= \frac{\sum_{j=1}^{T-B} \sum_{t=2}^B \sum_{k=1}^{t-1} u_{t+j} u_{k+j}}{B^{3/2} T^{1/2}} , \quad
	S_2 = \frac{(\rho-1)\sum_{j=1}^{T-B} \sum_{t=2}^B \sum_{k=1}^{t-1} u_{k+j} x_{t+j-1}}{B^{3/2} T^{1/2}} , \\
	S_3 &= \frac{(\rho-1) \sum_{j=1}^{T-B} \sum_{t=2}^B \sum_{k=1}^{t-1}  u_{t+j} x_{k+j-1}}{B^{3/2} T^{1/2}} , \ 
	S_4 = \frac{(\rho-1)^2 \sum_{j=1}^{T-B} \sum_{t=2}^B \sum_{k=1}^{t-1}  x_{t+j-1} x_{k+j-1}}{B^{3/2} T^{1/2}} .
\end{align*}
The first term is rearranged as
\begin{align*}
	S_1 = \sum_{t=1}^B \sum_{j=t+1}^{t+T-B} \sum_{k=1}^{t-1} \frac{u_{j} u_{k+j-t}}{B^{3/2} T^{1/2}} = \sum_{j=1}^T  \sum_{t \in \mathcal{I}_j} \sum_{k=1}^{t-1} \frac{u_{j} u_{j-k}}{B^{3/2} T^{1/2}} = \sum_{j=1}^{T} q_{j,T},
\end{align*}
which is a sum of elements of a martingale difference array.
For the second term, note that $E[\sum_{j=1}^{T-B} \sum_{t=2}^B \sum_{k=1}^{t-1} u_{k+j} x_0] = O(B^{3/2} T)$, which yields
\begin{align*}
	E[S_2]
	&= \frac{- c \sum_{j=1}^{T-B} \sum_{t=2}^B \sum_{k=1}^{t-1} \sum_{m=0}^{t+j-2} \rho^m E[u_{k+j} u_{t+j-1-m}]}{B^2 T}  + O(B^{-1/2}) \\
	&= \frac{- c \sum_{j=1}^{T-B} \sum_{t=2}^B \sum_{k=1}^{t-1} \rho^{t-k-1} E[u_{k+j}^2]}{B^2 T}  + O(B^{-1/2}) \\
	&= \frac{- c \sum_{j=1}^{T-B} \sum_{k=1}^{t-1} (B-k) E[u_{k+j}^2]}{B^2 T}  + O(B^{-1/2})
	= - \frac{c}{2} \int_0^1 \sigma^2(r) \dd r + o(1),
\end{align*}
and
\begin{align*}
	&E[S_2^2] = \frac{c^2 E [ ( \sum_{t=2}^B \sum_{k=1}^{t-1} \sum_{j=1}^{T-B} \sum_{m=0}^{t+j-2} \rho^m u_{k+j} u_{t+j-1-m} )^2 ] }{B^4 T^2} + O(B^{-1})  \\
	&= \frac{c^2 ( \sum_{t=2}^B \sum_{k=1}^{t-1} \sum_{j=1}^{T-B} \rho^{t-k-1} \sigma^2_{k+j} )^2}{B^4 T^2}  + O(B^{-1})
	= \frac{c^2}{4} \Big( \int_0^1 \sigma^2(r) \dd r \Big)^2 + o(1).
\end{align*}
Hence $Var[S_2] = o(1)$, and $S_2 = - c/2 \int_0^1 \sigma^2(r) \dd r + o_P(1)$.
Let $C = \max\{ C_1, C_2, 1 \} < \infty$, where $E[u_t^2] < C_1$, and $E[u_t^4] < C_2^2$, for all $t \in \mathbb{N}$.
Furthermore, let 
\begin{align*}
	\tilde S_3 &= \frac{(\rho-1) \sum_{j=1}^{T-B} \sum_{t=2}^B \sum_{k=1}^{t-1}  u_{t+j} (x_{k+j-1}-x_0)}{B^{3/2} T^{1/2}}, \\
	\tilde S_4 &= \frac{(\rho-1)^2 \sum_{j=1}^{T-B} \sum_{t=2}^B \sum_{k=1}^{t-1}  (x_{t+j-1}-x_0) (x_{k+j-1} - x_0)}{B^{3/2} T^{1/2}}.
\end{align*}
Then, $E[|S_3 - \tilde S_3|] = O(T^{-1/2})$, since $E[|T^{-1/2} \sum_{j=1}^{T-B} u_{t+j} x_0 |] = O(1)$, and
\begin{align*}
	E[|\tilde S_3|]  
	&\leq \frac{(1-\rho) \sum_{t=2}^B \sum_{k=1}^{t-1} \sqrt{E[(\sum_{j=1}^{T-B} \sum_{m=0}^{k+j-2} \rho^m u_{t+j} u_{k+j-1-m})^2]} }{B^{3/2} T^{1/2}} \\
	&\leq \frac{(1-\rho) C B^{1/2} \sum_{m=0}^T \rho^m }{T^{1/2}} 
	= O(B^{1/2} T^{-1/2}).
\end{align*}
Analogously, we have $E[|S_4 - \tilde S_4|] = O(T^{-1/2})$, and $E[|\tilde S_4|] = O(B^{1/2} T^{-1/2})$, which implies that $S_3 + S_4 = O_P(B^{1/2} T^{-1/2})$, and the assertion follows with $\mathcal W_T := -(S_1 + S_2 + S_3)/c$.

\textit{(b):}
Note that by mathematical induction on $n$, the identity $\sum_{t=2}^{n} \sum_{k=1}^{t-1} a_{k} = \sum_{k=1}^{n-1} (n -k) a_{k}$	
holds true for any sequence $(a_{t})_{t \in \mathbb{N}}$. 
The index set $\mathcal{I}_j$ can be expressed as
\begin{align*}
	\mathcal{I}_j = \begin{cases}   
	\{ t \in \mathbb{N}: \ 2 \leq t \leq j-1 \} & \text{if} \ j \in [1,B], \\
	\{ t \in \mathbb{N}: \ 2 \leq t \leq B \} & \text{if} \ j \in [B+1,T-B], \\
	\{ t \in \mathbb{N}: \ j+B-T \leq t \leq B \} & \text{if} \ j \in [T-B+1,T].
	\end{cases}
\end{align*}
For $j \in [1,B]$, it follows that
\begin{align}
	B^{3/2}T^{1/2} q_{j,T} = \sum_{t=2}^{j-1} \sum_{k=1}^{t-1} u_j u_{j-k} = \sum_{k=1}^{j-1} (j-1-k) u_j u_{j-k} = \sum_{k=1}^{j-2} k u_j u_{k+1}, \label{eq:mdvariance2}
\end{align}
and, analogously, if $j \in [B+1, T-B]$, we obtain
\begin{align}
	B^{3/2}T^{1/2} q_{j,T} = \sum_{t=2}^{B} \sum_{k=1}^{t-1} u_j u_{j-k} = \sum_{k=1}^{B} (B-k) u_{j-k} = \sum_{k=1}^{B-1} k u_j u_{j-B+k}.		\label{eq:mdvariance}
\end{align}
Let $i:=j+B-T$. If $j \in [T-B+1, T]$, or, equivalently, if $i \in [1,B]$, we have
\begin{align}
	&B^{3/2}T^{1/2} q_{i,T} 
	=  \sum_{t=i}^{B} \sum_{k=1}^{t-1} u_j u_{j-k} 
	= \sum_{t=2}^{B} \sum_{k=1}^{t-1} u_j u_{j-k} -  \sum_{t=2}^{i-1} \sum_{k=1}^{t-1} u_j u_{j-k} \nonumber \\
	&= \sum_{k=1}^{B} (B-k) u_j u_{j-k} - \sum_{k=1}^{i-1} (i-1-k) u_j u_{j-k}
	= \sum_{k=1}^{B-1} k u_j u_{j-B+k} - \sum_{k=1}^{i-2} k u_j u_{T-B+k+1}. \label{eq:mdvariance3}
\end{align}
Then,
\begin{align*}
Var\bigg[ \sum_{j=1}^{T} q_{j,T} \bigg] = \sum_{j=B+1}^{T-B} E[q_{j,T}^2] + o(1) =  \frac{1}{B^3 T} \sum_{j=B+1}^{T-B} \sum_{k=1}^{B-1} k^2 \sigma^2_j \sigma^2_{j-B+k} + o(1) = \Theta(1),
\end{align*}
and the first part of (b) has been shown.
For the second part, we decompose the denominator statistic into $\mathcal X_{2,T} = S_5 + S_6 + S_7$, where
\begin{align*}
	S_5 &= \frac{\sum_{j=1}^{T-B} \sum_{t=2}^B ( \sum_{k=1}^{t-1} u_{j+k} )^2}{B^2 T}, \quad
	S_6 = \frac{2(\rho-1)\sum_{j=1}^{T-B} \sum_{t=2}^B \sum_{k=1}^{t-1} \sum_{l=1}^{t-1} x_{j+k-1} u_{j+l}}{B^2 T}, \\
	S_7 &= \frac{(\rho-1)^2\sum_{j=1}^{T-B} \sum_{t=2}^B ( \sum_{k=1}^{t-1} x_{j+k-1} )^2}{B^2 T}.
\end{align*}
The first term satisfies 
\begin{align*}
	E[S_5] = \frac{1}{B^2 T}  \sum_{j=1}^{T-B} \sum_{t=2}^B \sum_{k=1}^{t-1} \sigma_{j+k}^2, \quad
	E[S_5^2] = \frac{1}{B^4 T^2} \Big(  \sum_{j=1}^{T-B} \sum_{t=2}^B \sum_{k=1}^{t-1} \sigma_{j+k}^2 \Big)^2 + \Theta(B T^{-1}),
\end{align*}
which yields $Var[S_5] = \Theta(B T^{-1})$.
Analogously to the result for $S_3$ and $S_4$, we obtain that
$E[|S_6|] + E[|S_7|] = O(B^{1/2} T^{-1/2})$,
which implies that $Var[\mathcal X_{2,T}] = \Theta(B T^{-1})$.

\textit{(c):}
With a constant error variance, equations \eqref{eq:mdvariance2}--\eqref{eq:mdvariance3} yield
\begin{align*}
	B^3 T \cdot E\big [q_{j,T}^2\big] =
	\begin{cases}
		\sigma^4 \sum_{k=1}^{j-2}  k^2 & \text{if} \ j \in [1,B], \\
		\sigma^4 \sum_{k=1}^{B-1}  k^2 & \text{if} \ j \in [B+1,T-B], \\
		\sigma^4 ( \sum_{k=1}^{B-1}  k^2  +  \sum_{k=1}^{i-2} (k^2 - 2k(B-k)))  & \text{if} \ j \in [T-B+1,T].
	\end{cases} 
\end{align*}
Combining all cases and applying the Gaussian summation formulas yields
\begin{align}
	Var[\mathcal X_{1,T}] 
	&= \sum_{j=1}^T E[q_{j,T}^2] 
	= \frac{\sigma^4}{B^3 T} \Big[ (T-B) \sum_{k=1}^{B-1} k^2 + \sum_{j=1}^B \sum_{k=1}^{j-2} [4k^2 - 2 Bk] \Big] \nonumber \\
	&= \sigma^4 \frac{(T-B)(B-1)(2B-1) - 2(B-1)(B-2)}{6 B^2 T}, \label{eq:auxhomosk1}
\end{align}
since $c=0$.
For the denominator, we have $S_6 = S_7 = 0$, since $c = 0$.
Then,
\begin{align}
	E[\mathcal X_{2,T}] = E[S_5] = \frac{1}{B^2 T}  \sum_{j=1}^{T-B} \sum_{k=1}^{B-1} (B-k) \sigma^2 = \sigma^2 \frac{(T-B)(B-1)}{2BT}, \label{eq:auxhomosk2}
\end{align}
and the assertion follows with equations \eqref{eq:auxhomosk1} and \eqref{eq:auxhomosk2}.
} \fi

\if0\supplement
{
  	The proof is available in the supplementary material.
} \fi

\subsection{Proof of Theorem \ref{thm_asy1_sb} }

From Lemma \ref{lem_NumDen}(a), it follows that $E[q_{j,T}^2] = O(T^{-1})$ for any $j \leq T$, which implies that $Var[\sum_{j=1}^T q_{j,T}] = \sum_{j=B+1}^{T-B} E[q_{j,T}^2] + o(1)$. 
The identity $\sum_{t=2}^{n} \sum_{k=1}^{t-1} a_{k} = \sum_{k=1}^{n-1} (n -k) a_{k}$	
holds true for any sequence $(a_{t})_{t \in \mathbb{N}}$, which follows by induction on $n$.
Then, for $B+1 \leq j \leq T-B$,
\begin{align*}
	B^{3/2}T^{1/2} q_{j,T} = \sum_{t=2}^{B} \sum_{k=1}^{t-1} u_j u_{j-k} = \sum_{k=1}^{B} (B-k) u_{j-k} = \sum_{k=1}^{B-1} k u_j u_{j-B+k},
\end{align*}
which yields 
\begin{align*}
	&Var\Big[\sum_{j=1}^T q_{j,T}\Big] 
	= \frac{\sum_{j=B+1}^{T-B} \sum_{k=1}^{B-1} k^2 E[u_j^2] E[u_{j-B+k}^2]}{B^3T}  + o(1) \\
	&= \int_{\frac{B}{T}}^{\frac{T-B}{T}} \int_0^1 s^2 \sigma^2(r) \sigma^2(\tfrac{j- \lfloor (1-s) B \rfloor}{T}) \dd s \dd r + o(1)  
	= \int_0^1 \int_0^1 s^2 \sigma^4(r) \dd s \dd r + o(1) \\
	&= \frac{1}{3} \int_0^1 \sigma^4(r) \dd r + o(1).
\end{align*}
Moreover, we have $\max_{1 \leq j \leq T} E[q_{j,T}^2] = o(1)$, and Jensen's and Markov's inequalities yield $\max_{1 \leq j \leq T} |q_{j,T}| = o_P(1)$.

Since $\{q_{j,T}\}$ is a martingale difference array, we can apply the central limit theorem from Theorem 24.3 in \cite{davidson1994}, which implies that $\sum_{j=1}^T q_{j,T}/\sqrt{Var[\sum_{j=1}^T q_{j,T}]} \Dlim \mathcal{N}(0,1)$,
as $T \to \infty$.
Furthermore, from Lemma \ref{lem_NumDen}, $E[\mathcal X_{1,T}] = - c/2 \int_0^1 \sigma^2(r) \dd r + o(1)$,
and the first statement follows from Lemma \ref{lem_blockfilter1}.
For the second statement, note that,
\begin{align*}
	&E [\mathcal X_{2,T}] = \frac{1}{B^2 T} \sum_{j=1}^{T-B} \sum_{t=2}^B E \bigg[ \Big( \sum_{k=1}^{t-1} \Delta x_{j+k} \Big)^2 \bigg]
	= \frac{1}{B^2 T} \sum_{j=1}^{T-B} \sum_{t=2}^B E \bigg[ \Big( \sum_{k=1}^{t-1} u_{j+k} + \phi x_{j+k-1} \Big)^2 \bigg] \\
	&= \frac{1}{B^2 T} \sum_{j=1}^{T-B} \sum_{t=2}^B E \bigg[ \Big( \sum_{k=1}^{t-1} u_{j+k} \Big)^2 \bigg] + o(1) 
	= \frac{1}{B^2 T}  \sum_{j=1}^{T-B} \sum_{t=2}^B \sum_{k=1}^{t-1} \sigma_{j+k}^2 + o(1) \\
	&= \frac{1}{B^2 T} \sum_{j=1}^{T-B} \sum_{k=1}^{B-1} (B-k) \sigma^2(\tfrac{j+k}{T}) + o(1)
	= \int_0^{\frac{T-B}{T}} \int_0^1 (1-s) \sigma^2(r+s\tfrac{B}{T}) \dd s \dd r + o(1) \\
	&= \int_0^{1} \int_0^1 (1-s) \sigma^2(r) \dd s \dd r + o(1) 
	= \frac{1}{2} \int_0^1 \sigma^2(r) \dd r + o(1).
\end{align*}
Furthermore, from Lemma 2, $Var[\mathcal X_{2,T}] = o(1)$, and the assertion follows by Chebyshev's inequality together with Lemma \ref{lem_blockfilter1}.

\subsection{Proof of Theorem \ref{thm_asy1_fb}}

Let $X_T(r) = T^{-{1/2}} \sum_{k=1}^{\rT} u_k$ and $Y_T(r) = T^{-{1/2}} x_{\rT}$ for $r \geq 0$.
From Lemmas 1 and 2 in \cite{cavaliere2005}, it follows that $X_T \Rightarrow  \ol \sigma  W_\eta$, where $\ol \sigma^2 = \int_0^1 \sigma^2(r) \dd r$ denotes the average variance.
For notational convenience, we set $u_0 = x_0$.
Note that a Taylor expansion around 0 yields $e^{-x} = 1 - x + o(x)$, which implies that $\rho = 1 - c/\sqrt{BT} = \exp ( - c/\sqrt{BT} ) + o(1/\sqrt{BT})$.
Then, with the continuous mapping theorem, we obtain
	\begin{align}
		&\frac{1}{\ol \sigma \sqrt{T}} x_{\rT} 
		=  \sum_{k=0}^{\rT} \rho^{\rT - k} \frac{u_k}{\ol \sigma \sqrt{T}} 
		= \sum_{k=0}^{\rT}  e^{-(\rT - k)c/\sqrt{BT}} \frac{u_k}{\ol \sigma \sqrt{T}}  + o_P(1) \nonumber \\
		&= \int_0^r e^{-(r-s)c/b} dX_T(s) + o_P(1) 
		\Rightarrow  \int_0^r e^{-(r-s)c/b} dW_{\eta}(s) = J_{c,b,\eta}(r), \label{eq:FCLT2aux}
	\end{align}
which yields $Y_T \Rightarrow \ol \sigma J_{c,b,\eta}$.
We rewrite
\begin{align*}
	&\Delta x_{t+j} x_{t+j-1} = \frac{\Delta x_{t+j}(x_{t+j-1} + x_{t+j} - \Delta x_{t+j})}{2} \\
	 &= \frac{(x_{t+j} - x_{t+j-1})(x_{t+j} + x_{t+j-1}) - (\Delta x_{t+j})^2}{2}
	 = \frac{x_{t+j}^2 - x_{t+j-1}^2 - (\Delta x_{t+j})^2}{2}
\end{align*}
such that
\begin{align*}
	&\sum_{t=2}^B \Delta x_{t+j} (x_{t+j-1} - x_j) 
	= \sum_{t=1}^B \frac{x_{t+j}^2 - x_{t+j-1}^2 - (\Delta x_{t+j})^2}{2} - \Delta x_{t+j} x_j \\
	&= \frac{1}{2} (x_{j+B}^2 - x_j^2) - (x_{j+B} x_j - x_j^2) - \frac{1}{2} \sum_{t=1}^B (\Delta x_{t+j})^2
	= \frac{(x_{j+B} - x_j)^2}{2} - \frac{1}{2} \sum_{t=1}^B (\Delta x_{t+j})^2.
\end{align*}
Then, with Lemma \ref{lem_blockfilter1},
\begin{align*}
	\mathcal Y_{1,T} &= \mathcal X_{1,T} + o_P(1) = \frac{\sum_{j=1}^{T-B} (x_{B+j} - x_j)^2 - \sum_{j=1}^{T-B} \sum_{t=1}^B (\Delta x_{t+j})^2}{2B^{3/2} T^{1/2}} \\
	&= \frac{\int_0^{1-b} (Y_T(b+r) - Y_T(r))^2 \dd r - \frac{1}{T^2}\sum_{j=1}^{T-B} \sum_{t=1}^B (\Delta x_{t+j})^2 }{2 b^{3/2}} + o_P(1).
\end{align*}
From $\Delta x_t = u_t$, it follows that
\begin{align*}
	E \bigg[ \frac{1}{T^2} \sum_{j=1}^{T-B} \sum_{t=1}^B (\Delta x_{t+j})^2 \bigg]
	= \frac{1}{T^2} \sum_{j=1}^{T-B} \sum_{t=1}^B E[u_{t+j}^2]
	= b(1-b) \int_0^1 \sigma^2(r) \dd r + o(1),
\end{align*}
which implies that
\begin{align}
	\mathcal Y_{1,T} = \frac{\int_0^{1-b} (Y_T(b+r) - Y_T(r))^2 \dd r - b(1-b) \int_0^1 \sigma^2(r) \dd r}{2 b^{3/2}} + o_P(1).	\label{f_Nfb}
\end{align}
Furthermore, Lemma \ref{lem_blockfilter1} yields
\begin{align}
	\mathcal Y_{2,T} = \mathcal X_{2,T} + o_P(1) = \frac{1}{b^2} \int_0^{1-b} \int_{r}^{b+r} (Y_T(s) - Y_T(r))^2 \dd s \dd r + o_P(1). \label{f_Dfb}
\end{align}
The assertion follows from equation \eqref{eq:FCLT2aux}, together with the continuous mapping theorem.

\subsection{Proof of Lemma \ref{lem_errorvariance} }

Since $(1 - \hat{\rho}) = O_P(B^{-1/2} T^{-1/2})$ and $x_t = O_P(T^{1/2})$, the residuals satisfy
\begin{align*}
	\hat{u}_t &= y_t - \hat{\rho} y_{t-1} 
	= \Delta y_t + (1-\hat \rho) y_{t-1} \\
	&= \Delta d_t + u_t + (\rho - 1) x_{t-1} + (1 - \hat \rho) y_{t-1} 
	= u_t + O_P(B^{-1/2})
\end{align*}
and $ \ol{ \hat u} = O_P(T^{-1/2})$. 
Then, for any $s \in [0,1]$,
\begin{align}
	\frac{1}{T} \sum_{j=1}^{\sT} (\hat{u}_j - \ol{ \hat u})^2 = \frac{1}{T} \sum_{j=1}^{\sT} u_{j}^2 + O_P(B^{-1/2}) = \int_0^s \sigma^2(r) \dd r + o_P(1),		\label{eq:Lem3aux1}
\end{align}
and (a) follows with $s=1$.
Furthermore, by Slutsky's theorem, $\hat{\eta}(s) = \eta(s) + o_P(1)$ holds pointwise for all $s \in [0,1]$.
Then, (b) follows by Dini's theorem since both $\hat{\eta}(s)$ and $\eta(s)$ are continuous, monotone, and bounded.
For (c), note that
\begin{align}
	\frac{1}{T-B} \sum_{j=1}^{T-B} \Big(\hat{u}_{j+t} - \frac{1}{B} \sum_{k=1}^B \hat{u}_{j+k} \Big)^2  =  \frac{1}{T-B} \sum_{j=1}^{T-B} u_{j+t}^2 + O_P(B^{-1/2}),	\label{eq:Lem3aux2}
\end{align}
for any $t=1, \ldots, B$.
Equations \eqref{eq:Lem3aux1} and \eqref{eq:Lem3aux2} yield
\begin{align*}
	&\frac{1}{(T-B)B} \sum_{j=1}^{T-B} \sum_{t=1}^B  \Big(\hat{u}_{j+t} - \frac{1}{B} \sum_{k=1}^B \hat{u}_{j+k} \Big)^2
	= \int_0^1 \sigma^2(r) \dd r + o_P(1), \\
	& \frac{1}{(T-B)B} \sum_{j=1}^{T-B} \sum_{t=1}^B ( \hat{u}_{j+1} - \ol{\hat u} )^2 \Big(\hat{u}_{j+t} - \frac{1}{B} \sum_{k=1}^B \hat{u}_{j+k} \Big)^2
	= \int_0^1 \sigma^4(r) \dd r + o_P(1),
\end{align*}
as $B,T \to \infty$ and $B/T \to 0$, and the result follows by Slutsky's theorem.

\subsection{Proof of Theorem \ref{thm_tstatistics} } 

Note that $v_T \to \sqrt{2/3}$, and $\hat \kappa v_T \sqrt{\mathcal Y_{2,T}} \overset{p}{\longrightarrow} \sqrt{\int_0^1 \sigma^4(r) \dd r / 3}$, which follows from Theorem \ref{thm_asy1_sb} and Lemma \ref{lem_errorvariance}.
Then, (a) follows together with Slutsky's theorem.
For (b), let $\tilde{x}_{\rT} = x_{\lfloor \hat{\eta}^{-1}(r) T \rfloor }$ and $\tilde{u}_{\rT} = u_{\lfloor \hat{\eta}^{-1}(r) T \rfloor }$.
Furthermore, let $\widetilde X_T(r) = T^{-1/2} \sum_{k=1}^{\rT} \tilde u_k$ and $\widetilde Y_T(r) = T^{-1/2} \tilde x_{\rT}$.
Theorem 1 in \cite{cavaliere2008} states that $\widetilde X_T \Rightarrow \ol \sigma W$, where $\ol \sigma^2 = \int_0^1 \sigma^2(r) \dd r$, and, analogously to \eqref{eq:FCLT2aux}, it follows that $\widetilde Y_T \Rightarrow J_{c,b}$.
Following equations \eqref{f_Nfb} and \eqref{f_Dfb}, we obtain
\begin{align*}
	\tau\text{-FB} = \frac{ ( \int_0^{1-b} ( \widetilde Y(b+r) - \widetilde Y(r) )^2 \dd r - b(1-b) \ol \sigma^2 ) / (2 b^{3/2}) }{\sqrt{ \ol \sigma \int_0^{1-b} \int_r^{b+r} ( \widetilde Y(s) - \widetilde Y(r) )^2 \dd s \dd r / b^2} } + o_P(1),
\end{align*}
and the assertion follows with the continuous mapping theorem and Slutsky's theorem.

\subsection{Proof of Lemma \ref{lem_prewhitenconsistency} }

\if1\supplement
{
  Let $\hat \gamma$ be the OLS estimator of $\Delta y_t$ on $z_t = (\Delta y_{t-1}, \ldots, \Delta y_{t-p_T}, T^{-1/2} y_{t-1})'$ for $t=p_T+1, \ldots, T$, which yields
$\hat \theta_i = e_i' \hat \gamma$, where $e_i$ is the $i$-th unit vector.
Let $f_t = (\Delta d_{t-1}, \ldots, \Delta d_{t-p_T}, T^{-1/2} d_{t-1})'$, $g_t = (x_{t-1}, \ldots, x_{t-p_T},0)'$, and $w_t = (u_{t-1}, \ldots, u_{t-p_T}, T^{-1/2} x_{t-1})'$.
Then, $z_t = f_t + \phi g_t + w_t$.
From equation \eqref{eq:augmentedregression}, we have $\Delta x_t = w_t'\beta + \epsilon_{p_T,t}$, where $\beta = (\theta_1, \ldots, \theta_{p_T}, T^{1/2} \phi)'$, and $\theta_i = e_i'\beta$ for all $i=1, \ldots, p_T$.
Consequently,
	\begin{align*}
		\Delta y_t = \Delta d_t + \Delta x_t = \Delta d_t + w_t'\beta + \epsilon_{p_T,t} = z_t'\beta + a_t,
	\end{align*}
	where $a_t = \epsilon_t + \sum_{i=p_T+1}^\infty \theta_i u_{t-i} + \Delta d_t - (f_t + \phi g_t)'\beta$.
	Then,
\begin{align*}
	\hat \gamma = \beta + \bigg(\sum_{t=p_T+1}^T z_t z_t' \bigg)^{-1} \sum_{t=p_T+1}^T z_t a_t.
\end{align*}	
Let $\|\cdot \|_1$ denote the $L_1$ vetor norm and its induced matrix norm, which is the maximum absolute column sum norm.
Then,
\begin{align*}
\bigg| \sum_{i=1}^{p_T} (\hat \theta_i - \theta_i) \bigg|
\leq p_T \bigg\| \bigg( \sum_{t=p_T+1}^T z_t z_t' \bigg)^{-1} z_t a_t \bigg\|_1
\leq \bigg\| \bigg( \frac{1}{T} \sum_{t=p_T+1}^T z_t z_t' \bigg)^{-1} \bigg\|_1 \cdot \bigg\| \frac{p_T}{T} \sum_{t=p_T+1}^T z_t a_t \bigg\|_1.
\end{align*}
Followwing the FCLT under time-varying variance (see Lemma 1 in \citealt{cavaliere2005}), Lemma 3 in \cite{berk1974}, and Lemma 3.2 in \cite{chang2002}, the eigenvalues uf $T^{-1} \sum_{t=p_T+1}^T z_t z_t'$ are bounded from above and below, which implies that
$\| (T^{-1} \sum_{t=p_T+1}^T z_t z_t' )^{-1} \|_1 = O_P(1)$.
It remains to show that $\| T^{-1} \sum_{t=p_T+1}^T z_t a_t \|_1 = O_P(B^{-1/2})$.
By the triangle inequality, we have $\| T^{-1} \sum_{t=p_T+1}^T z_t a_t \|_1 \leq A_1 + A_2 + A_3$, where
\begin{align*}
	A_1 &= \bigg\| \frac{1}{T} \sum_{t=p_T+1}^T (f_t + \phi g_t) a_t \bigg\|_1, \quad
	A_2 = \bigg\| \frac{1}{T} \sum_{t=p_T+1}^T w_t (\Delta d_t - (f_t + \phi g_t)'\beta) \bigg\|_1, \\
	A_3 &= \bigg\| \frac{1}{T} \sum_{t=p_T+1}^T w_t (\epsilon_t + \sum_{i=p_T+1}^\infty \theta_i u_{t-i}) \bigg\|_1.
\end{align*}
Note that $E[\| f_t + \phi g_t \|_1^2] = O(B^{-1})$, $E[\| w_t \|_1^2] = O(1)$, $E[(\Delta d_t - (f_t + \phi g_t)' \beta)^2] = O(B^{-1})$, and $E[(\epsilon_t + \sum_{i=p_t+1}^\infty \theta_i u_{t-i})^2] = O(1)$.
Then, by the Cauchy Schwarz inequality it follows that $E[|A_1 + A_2|] = O(B^{-1/2})$.
For $A_3$ note that $w_t (\epsilon_t + \sum_{i=p_T+1}^\infty \theta_i u_{t-i})$ is a martingale difference sequence, which yields $A_3 = O_P(T^{-1/2})$
Hence, $\| T^{-1} \sum_{t=p_T+1}^T z_t a_t \|_1 = O_P(B^{-1/2})$, and the assertion follows.

} \fi

\if0\supplement
{
  	The proof is available in the supplementary material.
} \fi
	
\subsection{Proof of Lemma \ref{lem_blockfilter4}}

\if1\supplement
{
  	The proof is split in two parts.
	In the first part, we show that
$\mathcal Y_{1,T}^* - \hat{\mathcal Y}_{1,T}^* = O_P(p_T B^{-1/2})$ and $ \mathcal Y_{2,T}^* - \hat{\mathcal Y}_{2,T}^* = O_P(p_T T^{-1/2})$, where 
\begin{align*}
	\mathcal Y_{1,T}^* = \frac{1}{B^{3/2} T^{1/2}} \sum_{j=1}^{T-B} \sum_{t=2}^B \Delta y^*_{t+j} (y^*_{t+j-1} - y^*_j), \quad 
	\mathcal Y_{2,T}^* = \frac{1}{B^2 T} \sum_{j=1}^{T-B} \sum_{t=2}^B (y^*_{t+j-1} - y^*_j)^2,
\end{align*}
and $y_t^* = \theta(L) y_t$.
We have
\begin{align*}
	\mathcal Y_{1,T}^* - \hat{\mathcal Y}_{1,T}^* &= \frac{1}{B^{3/2} T^{1/2}} \sum_{j=1}^{T-B} \sum_{t=2}^B \sum_{k=1}^{t-1} \big( \Delta y_{t+j}^* \Delta y_{k+j}^* - \Delta \hat y_{t+j}^* \Delta \hat y_{k+j}^* \big), \\
	\mathcal Y_{2,T}^* - \hat{\mathcal Y}_{2,T}^* &= \frac{1}{B^2 T} \sum_{j=1}^{T-B} \sum_{t=2}^B \sum_{k=1}^{t-1} \Big[ \big( \Delta y_{k+j}^* \big)^2 - \big( \Delta \hat y_{k+j} \big)^2 \Big].
\end{align*}
Let $\hat \theta(z) = 1- \sum_{i=1}^{p_T} \hat \theta_i z^i$, which yields $\hat y_t^* = \hat \theta(L) y_t$.
Furthermore, let $\theta^*(z) = \sum_{i=1}^\infty \theta_i^* z^i$, where
\begin{align*}
	\theta_i^* = \begin{cases}
	\theta_i - \hat \theta_i & \text{for} \ i \leq p_T, \\
	\theta_i & \text{for} \ i > p_T,
	\end{cases}
\end{align*}
which yields $\theta^*(L) y_t = y_t^* - \hat y_t^*$.
Let, for notational convenience, $\theta_0 = \hat \theta_0 = -1$.
Then,
\begin{align*}
	&\Delta y_{t+j}^* \Delta y_{k+j}^* - \Delta \hat y_{t+j}^* \Delta \hat y_{k+j}^*
	= (\Delta y_{t+j}^* - \Delta \hat y_{t+j}^*) \Delta y_{k+j}^* + \Delta \hat y_{t+j}^* (\Delta y_{k+j}^* - \Delta \hat y_{k+j}) \\
	&= \sum_{i=1}^\infty \theta_i^* \big[\Delta y_{t+j-i} \Delta y_{k+j}^* + \Delta \hat y_{t+j}^* \Delta y_{k+j-i}\big] \\
	&= - \sum_{i=1}^\infty \theta_i^* \Big( \sum_{m=0}^\infty \theta_m \Delta y_{t+j-i} \Delta y_{k+j-m} + \sum_{m=0}^{p_T} \hat \theta_m \Delta y_{t+j-m} \Delta y_{k+j-i} \Big),
\end{align*}
and
\begin{align*}
	&\big( \Delta y_{k+j}^* \big)^2 - \big( \Delta \hat y_{k+j} \big)^2
	= \big( \Delta y_{k+j}^* - \Delta \hat y_{k+j}^* \big)^2 + 2 \big( \Delta y_{k+j}^* - \Delta \hat y_{k+j}^* \big) \Delta \hat y_{j+j}^* \\
	&= \sum_{i,m=1}^\infty \theta_i^* \theta_m^* \Delta y_{k+j-i} \Delta y_{k+j-m} - 2 \sum_{i=1}^\infty \sum_{m=0}^{p_T} \theta_i^* \hat \theta_m \Delta y_{k+j-i} \Delta y_{k+j-m}.
\end{align*}
We decompose $\mathcal Y_{1,T}^* - \hat{\mathcal Y}_{1,T}^* = S_1 + S_2$ and $\mathcal Y_{2,T}^* - \hat{\mathcal Y}_{2,T}^* = S_3 + S_4$, where
\begin{align*}
	S_1 &= - \frac{1}{B^{3/2}} \sum_{t=2}^B \sum_{k=1}^{t-1} \sum_{i=1}^\infty \sum_{m=0}^\infty \theta_i^* \theta_m A_{t-i, k-m}, \quad
	S_2 = - \frac{1}{B^{3/2}} \sum_{t=2}^B \sum_{k=1}^{t-1} \sum_{i=1}^\infty \sum_{m=0}^{p_T} \theta_i^* \hat \theta_m A_{t-m, k-i}, \\
	S_3 &= \frac{1}{B^2 T^{1/2}} \sum_{t=2}^B \sum_{k=1}^{t-1} \sum_{i,m=1}^\infty \theta_i^* \theta_m^* A_{k-i, k-m}, \quad
	S_4 = -\frac{2}{B^2 T^{1/2}} \sum_{t=2}^B \sum_{k=1}^{t-1}\sum_{i=1}^\infty \sum_{m=0}^{p_T} \theta_i^* \hat \theta_m  A_{k-i, k-m},
\end{align*}
with $A_{r,s} = T^{-1/2} \sum_{j=1}^{T-B} \Delta y_{r+j} \Delta y_{s+j}$.
Note that
\begin{align*}
A_{r,s} &=
	\frac{1}{T^{1/2}} \sum_{j=1}^{T-B} \big( \Delta d_{r+j} + \phi x_{r+j-1} + u_{r+j} \big) \big( \Delta d_{s+j} + \phi x_{s+j-1} + u_{s+j} \big) \\
	&= \frac{1}{T^{1/2}} \sum_{j=1}^{T-B} \big( \phi x_{r+j-1} u_{s+j} + \phi u_{r+j} x_{s+j-1} + \phi^2 x_{r+j-1} x_{s+j-1} \big) + O_P(T^{-1/2}) \\
	&= O_P(B^{-1/2}),
\end{align*}
and, by Lemma \ref{lem_prewhitenconsistency}, $\sum_{i=1}^{p_T} \theta_i^* = O_P(p_T B^{-1/2})$.
It then follows that $S_1 + S_2 = O_P(p_T B^{-1/2})$, and $S_3 + S_4 = O_P(p_T B^{-1} T^{-1/2})$.
Hence, the first part of the proof is completed.

For the second part of the proof, note that $d_t^* = y_t^* - x_t^* = \theta(L) d_t = d^*(t/T)$ is Lipschitz continuous, since
	\begin{align*}
		|d^*(t/T) - d^*(s/T)| = \theta(L) (d_t - d_s) \leq C \Big(1 - \sum_{i=1}^\infty |\theta_i| \Big)  \Big| \frac{t-s}{T} \Big|,
	\end{align*}
	where $C (1 - \sum_{i=1}^\infty |\theta_i|) < \infty$. Hence, under Assumption \ref{ass_lrv}, the transformed statistics $\mathcal Y_{i,T}^*$ and $\mathcal X_{i,T}^*$, $i=1,2$, have the same properties as $\mathcal Y_{i,T}$ and $\mathcal X_{i,T}$, $i=1,2$, under Assumption \ref{ass:errors}.
	Thus, Lemma \ref{lem_blockfilter1} yields $\mathcal Y_{1,T}^* - \mathcal X_{1,T}^* = O_P(B^{-1/2})$ and $\mathcal Y_{2,T}^* - \mathcal X_{2,T}^* = O_P(T^{-1/2})$.
Finally, the triangle inequality implies that $\hat{\mathcal Y}_{1,T}^* - \mathcal X_{1,T}^* = O_P(p_T B^{-1/2})$ and $\hat{\mathcal Y}_{2,T}^* - \mathcal X_{2,T}^* = O_P(p_T B^{-1/2})$.
} \fi

\if0\supplement
{
  	The proof is available in the supplementary material.
} \fi

\subsection{Proof of Theorem \ref{thm_shortrun}}

Let, for notational convenience, $\theta_0 = \hat \theta_0 = -1$, and let $\hat \theta(z) = 1- \sum_{i=1}^{p_T} \hat \theta_i z^i$, which yields $\hat y_t^* = \hat \theta(L) y_t$. 
Let $\hat d_t^* = \hat \theta(L) d_t$ and $\hat x_t^* = \hat \theta(L) x_t$.
%
Analogously to the proof of Lemma \ref{lem_errorvariance}, we have
\begin{align*}
	\hat u_t^* = \hat y_t^* - \hat \rho^* \hat y_{t-1} = \Delta \hat y_t^* + (1- \hat \rho^*) \hat y_{t-1}^*
	= \Delta \hat x_t^* + O(B^{-1/2})
	= \epsilon_t + o_P(1).
\end{align*}
The consistencies of $\hat \sigma^{*2}$, $\hat \kappa^{*2}$, and $\hat \eta^*(s)$ follow from the fact that
\begin{align*}
	\frac{1}{T} \sum_{j=1}^{\sT} (\hat{u}_j^* - \ol{ \hat u^*})^2 = \frac{1}{T} \sum_{j=1}^{\sT} \epsilon_j^2 + o_P(1) = \int_0^s \sigma^2(r) \dd r + o_P(1), \quad s \in [0,1],
\end{align*}
and
\begin{align*}
	&\frac{1}{T-B} \sum_{j=1}^{T-B} \Big(\hat{u}_{j+t}^* - \frac{1}{B} \sum_{k=1}^B \hat{u}_{j+k}^* \Big)^2  =  \frac{1}{T-B} \sum_{j=1}^{T-B} \epsilon_{j+t}^2 + o_P(1), \\
	&\frac{1}{(T-B)B} \sum_{j=1}^{T-B} \sum_{t=1}^B  \Big(\hat{u}_{j+t}^* - \frac{1}{B} \sum_{k=1}^B \hat{u}_{j+k}^* \Big)^2
	= \int_0^1 \sigma^2(r) \dd r + o_P(1), \\
	& \frac{1}{(T-B)B} \sum_{j=1}^{T-B} \sum_{t=1}^B ( \hat{u}_{j+1}^* - \ol{\hat u^*} )^2 \Big(\hat{u}_{j+t}^* - \frac{1}{B} \sum_{k=1}^B \hat{u}_{j+k}^* \Big)^2
	= \int_0^1 \sigma^4(r) \dd r + o_P(1),
\end{align*}
where the last two equations hold true as $B/T \to 0$, analogously to Lemma \ref{lem_errorvariance}.

Finally, since the pre-whitened numerator and denominator statistics $(\mathcal{X}_{1,T}^*, \mathcal{X}_{2,T}^*)$ under Assumption \ref{ass_lrv} have the same properties as $(\mathcal{X}_{1,T}, \mathcal{X}_{2,T})$ under Assumption \ref{ass:errors}, the assertion follows with Lemma \ref{lem_blockfilter4} and the proof of Theorem \ref{thm_tstatistics}.

\end{document}